\documentclass[twocolumn,showpacs,preprintnumbers,amsmath,amssymb,superscriptaddress]{revtex4-1}

\usepackage{graphicx}
\usepackage{dcolumn}
\usepackage{amsmath,mathrsfs,amssymb,amsthm}
\usepackage{hhline}
\usepackage{wasysym,enumerate,color}
\usepackage{epstopdf}
\usepackage{verbatim}
\usepackage{hyperref}
\usepackage{color}
\usepackage{ulem} 

\renewcommand{\emph}[1]{\textit{#1}} 

\definecolor{darkgreen}{rgb}{0,0.5,0}
\definecolor{purple}{rgb}{0.6,0,0.5}
\definecolor{orange}{rgb}{1,0.5,0}
\definecolor{darkred}{rgb}{.7,0,0}
\definecolor{darkblue}{rgb}{0,0,.3}
\definecolor{grey}{rgb}{.6,.6,.6}
\definecolor{dimgreen}{rgb}{0.2,0.6,0.1}
\hypersetup{colorlinks,linkcolor=blue,urlcolor=blue,citecolor=blue}

\newcommand{\changes}[1]{{\color{black}{#1}}}

\newcommand{\backup}[1]{}
\newcommand{\hatnd}{\hat n_d}
\newcommand{\nd}{n_d}
\newcommand{\ndsigma}{n_{d\sigma}}
\newcommand{\ndup}{n_{d\uparrow}}
\newcommand{\nddown}{n_{d\downarrow}}
\newcommand{\hatndup}{\hat n_{d\uparrow}}
\newcommand{\hatnddown}{\hat n_{d\downarrow}}
\newcommand{\hatndsigma}{\hat n_{d \sigma}}
\newcommand{\md}{m_d}
\newcommand{\HFL}{H_{\rm FL}}
\newcommand{\imp}{{\rm imp}}
\newcommand{\bulk}{{\rm bulk}}
\newcommand{\Tk}{T_K}
\newcommand{\Estar}{E^\ast}
\newcommand{\nzero}{n^0}

\newcommand{\ezero}{{\varepsilon_0}}

\newcommand{\muzero}{{\mu_0}} 
 
\newcommand{\ed}{{\varepsilon_d}} 
\newcommand{\eprime}{{\varepsilon'}}
\newcommand{\sprime}{{\sigma'}}

\newcommand{\qqph}{\qquad \phantom{.}}

\newcommand{\Eq}[1]{Eq.~(\ref{#1})}

\newcommand{\Eqs}[1]{Eqs.~(\ref{#1})}

\usepackage{color}

\newcommand{\be}{\begin{equation}}
\newcommand{\ee}{\end{equation}}

\newcommand{\strongcoupling}{strong-coupling}

\begin{document}
\title{Fermi-liquid theory for the single-impurity Anderson model}
\author{Christophe Mora}
\affiliation{Laboratoire Pierre Aigrain, Ecole Normale Sup\'erieure, CNRS, UPMC,
Universit\'e Paris Diderot, 24 rue Lhomond, 75005 Paris, France}
\author{C\u at\u alin Pa\c scu Moca}
\affiliation{BME-MTA Exotic Quantum Phase Group, Institute of Physics, Budapest University of Technology and Economics,
H-1521 Budapest, Hungary}
\affiliation{Department of Physics, University of Oradea, 410087, Oradea, Romania}
\author{Jan von Delft}
\affiliation{Physics Department, Arnold Sommerfeld Center for Theoretical Physics and Center for NanoScience, Ludwig-Maximilians-Universit\"at M\"unchen, 80333 M\"unchen, Germany}
\author{Gergely Zar\'and}
\affiliation{BME-MTA Exotic Quantum Phase Group, Institute of Physics, Budapest University of Technology and Economics,
H-1521 Budapest, Hungary}
\date{\today}

\begin{abstract}
  We generalize Nozi\`eres' Fermi-liquid theory for the low-energy
  behavior of the Kondo model to that of the single-impurity Anderson
  model.  In addition to the electrons' phase shift at the Fermi
  energy, the low-energy Fermi-liquid theory is characterized by four
  Fermi-liquid parameters: the two given by Nozi\`eres
    that enter to first order in the excitation energy, and two
    additional ones that enter to second order and are needed away
    from particle-hole symmetry. We express all four parameters in
  terms of zero-temperature physical observables, namely the local
  charge and spin susceptibilities and their derivatives w.r.t.\ the
  local level position. We determine these in terms of the bare
  parameters of the Anderson model using Bethe Ansatz and Numerical
  Renormalization Group (NRG) calculations.  Our low-energy
  Fermi-liquid theory applies throughout the crossover from the
  \strongcoupling\ Kondo regime via the mixed-valence regime to the
  empty-orbital regime.  From the Fermi-liquid theory, we determine
  the conductance through a quantum dot symmetrically coupled to two
  leads in the regime of small magnetic field, low temperature and
  small bias voltage, and compute the coefficients of the $\sim B^2$,
  $\sim T^2$, and $\sim V^2$ terms \textit{exactly} in
    terms of the Fermi-liquid parameters.  The coefficients of $T^2$,
  $V^2$ \changes{and $B^2$} are found to change sign during the Kondo to empty-orbital
  crossover. The crossover
  becomes universal in the limit that the local interaction is much
  larger than the level width. For completeness, we
    also compute the shot noise and discuss the resulting Fano
    factor.
\end{abstract}


\maketitle

\section{Introduction and Summary}

\subsection{Introduction}

The single-impurity Anderson model, originally introduced to describe
d-level impurities such as Fe or Mn in metallic alloys
\cite{gruner1974,tsvelick1983,hewson1997kondo}, may well be one
of the most intensely studied models in condensed matter physics,
since it covers a rich variety of behaviors and non-perturbative
effects, including spin formation, mixed-valence
physics, and Kondo screening.  Indeed, various extensions of the
Anderson model underlie our understanding of correlated metals and
superconductors, Mott insulators~\cite{georges1996},
non-Fermi-liquid systems~\cite{cox1998}, and heavy fermion
materials~\cite{coleman2007}.

The Anderson model has also emerged as a standard tool to describe
Coulomb blockade in electron transport through quantum dot nanodevices
\cite{glazman2003,*glazman2004,chang2009}. Since quantum dots can
experimentally be probed under nonequilibrium conditions, this opened
a new chapter in the study of the Anderson model, involving its
properties in the context of nonequilibrium transport. This raised
novel questions, not relevant for impurities in bulk systems,
involving the behavior of the nonlinear conductance through a quantum
dot as a function of source-drain bias. To date, no exact results 
are available for the nonlinear conductance through a quantum dot 
described by an Anderson model away from its electron-hole symmetrical point. 

In the present paper, we fill this gap, albeit only at low energies,
by developing a Fermi-liquid (FL) theory for the low-energy behavior
of the asymmetric Anderson model. The theory is
  similar in spirit to the FL theory developed by Nozi\`eres for the
  Kondo model, but employs two additional FL parameters, whose form had not
  been established up to now. While these parameters do not influence quantities such as the 
  Wilson ratio, they are necessary to determine non-equilibrium transport properties such as shot noise or 
  the non-linear conductance discussed here.  We show how to express all FL
  parameters of our theory in terms of the zero-temperature,
equilibrium values of physical quantities such as the charge and spin
susceptibilities and the linear conductance.  Such a Fermi-liquid
theory is useful, because it offers an exact description of the
system's low-energy excitations, induced, e.g., by a small temperature
or a nonequilibrium steady-state transport due to a small source-drain
voltage. In this way, knowledge of ground state properties can be
elegantly used to make exact predictions about low-lying excitations.

\subsection{Anderson model basics}

In its simplest form, the Anderson model consists of a single spinful
interacting level of energy $\varepsilon_d$ and occupation $\hatnd =
  \hatndup + \hatnddown$, described by the simple Hamiltonian \be H_d
= \varepsilon_d \, \hatnd + \frac {U}{2} \hatnd^2 \;, \ee which is
coupled by a tunneling rate $2\Delta$ to the Fermi sea of spinful
conduction electrons.  In the presence of a local magnetic field,
  the level is Zeeman-split by an additional term $(\hatndup -
  \hatnddown)B/2$ (we use units where the Lande factor
times Bohr magneton give $g \mu_{\rm B} = 1$).  In the
non-equilibrium context of nano-devices, -- also discussed here, --
the level may be coupled to several leads characterized by 
different tunneling rates and Fermi energies.  As mentioned before, this simple
model exhibits a surprisingly rich behavior.  In particular, in the
limit of small $\Delta$ and a single electron on the level, i.e.\ an
average charge $\nd = \langle \hatnd \rangle \approx 1$, a local
magnetic moment is formed on the level. In this ``Kondo limit'', 
formally achieved for~\footnote{In fact, the Kondo limit can be formally
extended to the region $\varepsilon_d/U \in [-1,0]$ when  $U/\Delta \gg 1$.
In this limit, the potential scattering term, which breaks particle-hole
symmetry, vanishes~\cite{hewson1997kondo}.}
\begin{eqnarray}
\label{eq:Kondo-limit}
  \ed = -U/2,  \qquad   U/\Delta \gg 1   \; , 
\end{eqnarray}
the Anderson
model maps onto the Kondo model at small
energies~\cite{schrieffer1966} and accounts for the Kondo
effect~\cite{kondo1964,hewson1997kondo}, i.e.\ the dynamical
screening of the spin of this localized electron at low temperatures. 

Despite being the realm of strong correlations, the low-energy
structure of the screened Kondo state can be captured by simple means.
Following Wilson's solution of the Kondo model by the numerical
renormalization group~\cite{wilson1975}, Nozi\`eres realized that the
low temperature behavior of the Kondo model can be described as a
\emph{local Fermi liquid}, and can be understood in terms of weakly
interacting \emph{quasiparticles}. He formulated an effective
    Fermi-liquid theory for these, in terms of the phase shift that a
    quasipaticle incurs when scattering off the screened singlet
    \cite{nozieres1974}.  This phase shift, say $\delta_\sigma
    (\varepsilon, n_{\sigma'})$, depends not only on the kinetic energy
    $\varepsilon$ and spin $\sigma$ of the quasiparticle, but also on
    the entire distribution function $ n_{\sigma'} (\varepsilon')$ of the
    quasiparticles with which it interacts.  Nozi\`eres expanded this
    phase shift to leading order in $\varepsilon$ and the deviation
    $\delta n_{\sigma'} (\varepsilon')$ of the quasiparticle distribution
    function from its ground-state form, and viewed the two expansion
    coefficients as phenomenological parameters, $\alpha_1$ and
    $\phi_1$, called Fermi-liquid parameters. These parameters can be
    viewed as coupling constants in an effective Fermi-liquid
    Hamiltonian, which, when treated in the Hartree approximation,
    generates the phase shifts. The parameters $\alpha_1$ and $\phi_1$ can be
    expressed in terms of zero-temperature physical observables by
    exploiting the fact that the phase shifts determine, via the
    Friedel sum rule, the local charge and magnetization at zero
    temperature. In this way, both $\alpha_1$ and $\phi_1$ are found
    to be proportional to the zero-temperature impurity spin
    susceptibility, $\chi_s$, whose inverse defines the Kondo
    temperature, $T_K$, the characteristic low-energy scale of the
    Kondo model. 

Using the resulting  quasiparticle
Fermi-liquid (quasiparticle FL) theory,
Nozi\`eres~\cite{nozieres1974,*nozieres1974b,*nozieres1978} was able
to reconstruct all essential low temperature characteristics of the
Kondo model, such as the value of the anomalous Wilson ratio (the
dimensionless ratio of the impurity's contribution to the
susceptibility and to the linear specific heat coefficient),
$R=2$~(see Ref.~\cite{wilson1975}), or the quadratic temperature and magnetic
field dependence of the resistivity. 

Independently, Yamada and Yoshida developed a diagrammatic
Fermi-liquid
theory~\cite{yosida1970,*yamada1975a,*yosida1975,*yamada1975b}: they
reproduced the above-mentioned features within the Anderson model by
means of a perturbative approach and demonstrated by using Ward
identities that they hold up to infinite order in $U$.

Both the quasiparticle and the diagrammatic Fermi-liquid approaches
proved to be extremely useful. The diagrammatic FL approach has been
extended to orbitally degenerate versions of the Anderson
model~\cite{yoshimori1976,mihaly1978,horig2014,hanl2014}, see also
  the interaction between two impurities~\cite{schlottmann1980}, and
to out of equilibrium~\cite{oguri2001}, and led to the construction
of the renormalized perturbation
theory~\cite{hewson1993,*hewson1993b,*hewson1994,hewson1997kondo,hewson2001,*hewson2004,*hewson2006,*hewson2006b,*bauer2007,*edwards2011,*edwards2013}
(see also Ref.~\cite{streib2013}) and its application to various
extensions of the Anderson
model~\cite{fujii2010,sakano2011,sakano2011b}. Nozi\`eres'
quasiparticle FL approach has been widely used to study
non-equilibrium transport in correlated nano-structures described by
the Kondo model or generalizations
thereof~\cite{glazman2005,sela2006,golub2006,gogolin2006,mora2008,mora2009a,mora2009,vitushinsky2008}. In
particular, the effective Fermi-liquid Hamiltonian of the Kondo model
was used to calculate the leading dependence of the conductance on
temperature, bias voltage and magnetic field, and to determine the
coefficients of the leading $T^2/T_K$, $V^2/T_K^2$ and $B^2/T_K^2$
terms, say $c_T$, $c_V$ and $c_B$. These Fermi-liquid transport
coefficients turn out to be universal numbers, because for the Kondo
model the zero-energy phase shift, $\delta_0$, has a universal value,
$\delta_0 = \pi/2$.

Surprisingly, Nozi\`eres' quasiparticle Fermi-liquid theory has not
yet been extended to the case of the Anderson model (except for the
special case of electron-hole symmetry~\cite{sela2009}), although this
model has a Fermi-liquid ground state in all parameter
regimes~\cite{haldane1978,krishna1980,*krishna1980a}.  The reason has
probably been that such a theory requires additional Fermi-liquid
parameters, called $\phi_2$ and $\alpha_2$ below,
and no strategy was known to relate these to physical observables. In
this work, we fill this gap and develop a comprehensive Fermi-liquid
approach to the Anderson model, applicable also away from
particle-hole symmetry~\cite{munoz2013,merker2013}.  Our strategy is a
natural generalization of that used by Nozi\`eres for the Kondo model.
We develop an effective quasiparticle theory characterized by four
Fermi liquid parameters ($\alpha_1$,  $\phi_1$, $\alpha_2$  and
$\phi_2$), and use these to expand the phase shifts of the
quasiparticles systematically as a function of the quasiparticles'
energy and distribution. Using the Friedel sum rule, we express these
Fermi-liquid parameters in terms of four zero-temperature physical
parameters, namely the local charge and spin susceptibilities,
$\chi_c$ and $\chi_s$, and their derivatives $\chi'_c$ and $\chi'_s$
w.r.t.\ the local level position $\varepsilon_d$. We then use the
resulting Fermi-liquid Hamiltonian for the Anderson model to calculate
the conductance to quadratic order in temperature, bias voltage and
magnetic field, in a similar manner as for the Fermi-liquid
Hamiltonian for the Kondo model. However, the Fermi-liquid transport
coefficients $c_T$, $c_V$ and $c_B$ are no longer universal, but
depend on $\chi_c$, $\chi_s$, $\chi'_c$, $\chi'_s$ and the zero-energy
phase shift $\delta_0$, all of which are functions of $\varepsilon_d$.
For completeness, we also compute the current noise to
  third order in the voltage.  We calculate these functions
explicitly by using Bethe Ansatz and
NRG\cite{wilson1975,krishna1980,*krishna1980a}.  We thus obtain
explicit results for the $\varepsilon_d$ dependence of $c_T$, $c_V$,
$c_B$ and the current noise throughout the entire
crossover from the strong-coupling Kondo regime ($- U + \Delta
\lesssim \varepsilon_d \lesssim - \Delta$) via the mixed-valence
regime ($- \Delta \lesssim \varepsilon_d \lesssim \Delta$) to the
empty-orbital regime ($\varepsilon_d \gtrsim \Delta$).

\subsection{Summary and overview of main results}
\label{sec:summary}

In this subsection, we gather the main ideas of our approach and its main
results in the form of an executive summary. Details of their
derivation are presented in subsequent sections.

We shall focus on the quantum dot configuration 
\changes{connected symmetrically to
two lead reservoirs}. In this case, the level on the dot
couples only to the `symmetrical' combination of electronic states \changes{in
the leads}.  
Correspondingly, the Fermi-liquid theory can be constructed in
terms of quasiparticles in `even' and `odd' channels, $b$ and $a$,
respectively~\cite{mora2009}. Since the `odd' quasiparticles do not
hybridize with the $d$-level, the effective low-energy
Fermi-liquid Hamiltonian can be constructed solely from the
`even' quasiparticles, and is given to leading and subleading order by
 \begin{eqnarray}
\HFL & =&   \sum_\sigma \int_\varepsilon \, ( \varepsilon -\sigma B/2) \; b_{\varepsilon \sigma}^\dagger b_{\varepsilon \sigma} +H_\alpha + H_\phi +\dots
\label{eq:H_FL}
\\
H_\alpha&=& - \! \sum_\sigma \int_{\varepsilon_{1},\varepsilon_2} 
\!\!
 \left[ \frac{\alpha_1}{2\pi}  \bigl({\varepsilon_1 + \varepsilon_2}\bigr) + 
 \frac{\alpha_2}{4\pi}\bigl({\varepsilon_1 + \varepsilon_2}\bigr)^2 \right]  
\!  b_{\varepsilon_1 \sigma}^\dagger b_{\varepsilon_2 \sigma}
 \nonumber
\\
\nonumber
H_\phi&=&
  \int_{\varepsilon_{1},\dots,\varepsilon_4} 
\left[ \frac{\phi_1}{\pi} + \frac{\phi_2}{4\pi} 
(\sum_{i=1}^4 \varepsilon_i)  
\right]
: b_{\varepsilon_1 \uparrow}^\dagger  b_{\varepsilon_2 \uparrow}  b_{\varepsilon_3 \downarrow}^\dagger  b_{\varepsilon_4 \downarrow}  :,
\end{eqnarray}
where $B$ is the magnetic field.
Here $\alpha_1$, $\phi_1$, $\alpha_2$ and $\phi_2$ are the four
  Fermi-liquid parameters. The form of Eq.~\eqref{eq:H_FL} 
can be justified rigorously using Conformal Field Theory arguments as discussed in the Supplemental Material~\footnote{See Supplemental Material for a detailed discussion about some analytical and numerical calculations.}.  The
operators $b_{\varepsilon \sigma}^\dagger $ here create incoming
single-particle scattering states of kinetic energy
$\varepsilon$ and spin $\sigma$, and incorporate already the
zero-temperature phase shift $\delta_0$ experienced by electrons at
the Fermi energy, $\varepsilon=0$. The term $H_\alpha$ in this
expansion accounts for energy dependent elastic scattering, while the
terms in $H_\phi$ describe local interactions between the
quasiparticles. In the Kondo model, charge fluctuations are
suppressed, and the low-energy theory exhibits electron-hole symmetry
under the transformation $b_{\varepsilon \sigma}^\dagger
\leftrightarrow b_{-\varepsilon \sigma}$. In the presence of such
symmetry, the parameters $ \alpha_2$ and $ \phi_2$ must vanish,
since their presence would violate electron-hole
symmetry. Furthermore, as shown by
Nozi\`eres~\cite{nozieres1974,*nozieres1974b,*nozieres1978}, the
parameters $\alpha_1$ and $\phi_1$ are equal in the Kondo
model. Therefore the Kondo model's effective FL theory \eqref{eq:H_FL}
is characterized by a single Fermi-liquid scale, $\Estar$, defined
as 
\be {\Estar} \equiv \frac {\pi}{4 \alpha_1}\;,
\label{eq:Delta*}
\ee 
and identified as the Kondo temperature, $\Estar=T_K$. \changes{We use
units in which $k_B=1$}. In
contrast, in the generic Anderson model, three of the four
Fermi-liquid parameters are independent (more precisely, each of them
is a function of three variables, $\Delta$, and the dimensionless
ratios $\varepsilon_d/U$ and $\varepsilon_d/\Delta$), and therefore
the low-energy behavior cannot be characterized by a single
Fermi-liquid scale. Nevertheless, we shall still use
Eq.~\eqref{eq:Delta*} to define the characteristic energy scale
$\Estar$ and express physical quantities in terms of it.  We emphasize
that whereas the calculation of Nozi\`eres accounted only for local
spin excitations, our approach includes both spin and charge
fluctuations and allows us to capture the mixed-valence regime and
smoothly interpolate between the Kondo and Coulomb blockade regions.

To make use of the Fermi-liquid theory in its full power, we shall
determine the Fermi-liquid parameters in Eq.~\eqref{eq:H_FL} in terms
of the bare parameters of the Anderson model, $U$, $\varepsilon_d$, and
$\Delta$.  To this end, we shall first demonstrate that the four FL
parameters of the Anderson model are directly related to
zero-temperature physical observables, and can be expressed
solely in terms of the local charge ($\chi_c$) and spin ($\chi_s$)
susceptibilities of the Anderson model and their derivatives
($\chi_c'$ and $\chi_s'$) with respect to $\varepsilon_d$,
\begin{subequations}\label{fermic}
\begin{align}
&\frac{\alpha_1}\pi  =  \chi_s + \frac{\chi_c}{4} \;, 
\hspace{0.7cm}
\frac{\alpha_2}\pi    = -  \frac{3}{4} \chi_s' - \frac{\chi_c'}{16}\;,
\\
&
\frac {\phi_1}\pi  =  \chi_s - \frac{\chi_c}{4} \;, 
\hspace{0.7cm}
\frac{\phi_2} \pi =   - \chi_s' + \frac{\chi_c'}{4} \; .
\end{align} 
\end{subequations}
The expressions for $\alpha_1$ and $\phi_1$ were
  known~\cite{yamada1975a,yosida1975,hewson1993,hewson1997kondo} (see
  Sec. S-I in~\cite{Note2}) , those for $\alpha_2$ and $\phi_2$ are
   central  results of this work.  We then determine the FL
parameters from these relations, by computing the susceptibilities
$\chi_c(\varepsilon_d,\Delta,U)$ and $\chi_s(\varepsilon_d,\Delta,U)$
from NRG~\cite{wilson1975,krishna1980,*krishna1980a} and,
complementarily, by computing the Bethe Ansatz solution to the
Anderson model~\cite{kawakami1982ground,tsvelick1983a}.

\begin{figure}[tbhp]
\includegraphics[width=\columnwidth]{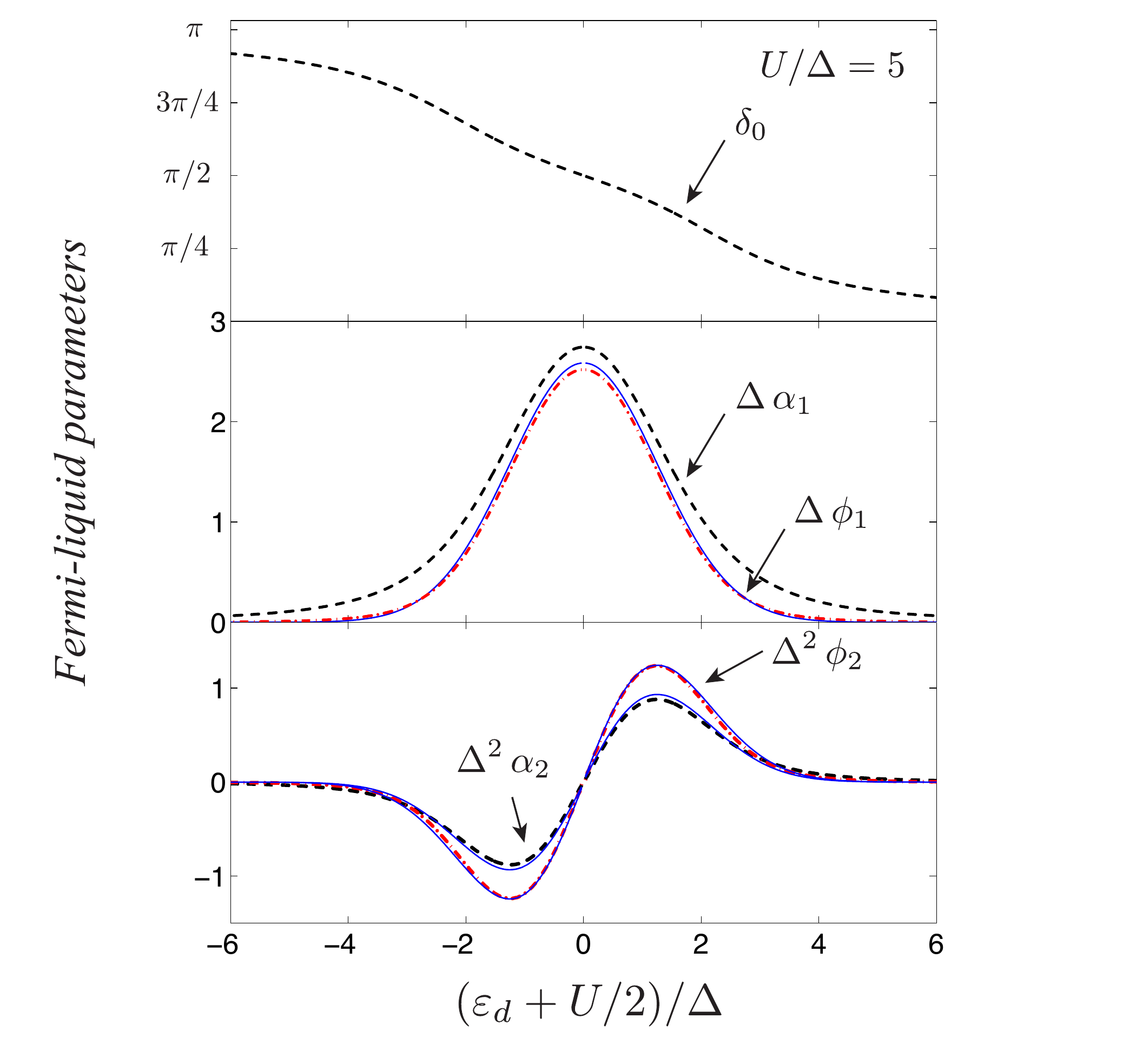}
\caption{\label{fig-fermicoeff}(Color online) Fermi-liquid parameters,
  $\alpha_{1,2}$ (dashed line) and $\phi_{1,2}$ (dash-dotted line), in
  units of $1/\Delta$ (or $1/\Delta^2$), as functions of
  $(\varepsilon_d+U/2)/\Delta$ for $U=5 \Delta$, calculated from
    \Eqs{fermic}, with the susceptibilities occurring therein
  extracted from the Bethe Ansatz computations.  Charge degeneracy
  occurs for $\varepsilon_d+U/2 \simeq 2.5 \Delta$. The thin
  continuous lines were computed using the analytical formulas,
  Eqs.~\eqref{asympto-kondo} and \eqref{asympto-kondo2}, valid in the
  Kondo regime. We also include the zero-energy phase shift $\delta_0$
  (dashed line) in the top panel, obtained from the  Friedel sum
  rule Eq.~\eqref{define-charge-magnetization} (at $B=0$) and the 
  Bethe Ansatz calculation of $n_d$.
 }
\end{figure}

Typical results of our computations are shown in
Fig.~\ref{fig-fermicoeff}, where we display the four Fermi-liquid
parameters for moderately strong interactions, $U/\Delta=5$, as a
function of the level's position. In agreement with the discussion
above, the parameters $\alpha_2$ and $\phi_2$ vanish at the
electron-hole symmetrical point, $\varepsilon_d=-U/2$, and are
antisymmetrical with respect to it, while the Fermi-liquid
parameters $\alpha_1$ and $\phi_1$ display a symmetrical
behavior. In the local-moment regime, $\langle \nd \rangle\approx 1$,
charge fluctuations are suppressed, and the charge susceptibility
$\chi_c$ can be neglected in the expression of the Fermi-liquid
parameters. Here we can derive an analytical approximation for them
[Eqs.~\eqref{asympto-kondo} and \eqref{asympto-kondo2}] by making use
of the Bethe Ansatz expression for the spin susceptibility in the
local-moment regime, $\chi_s\sim T_K^{-1}$.  Although
Eqs.~\eqref{asympto-kondo} and \eqref{asympto-kondo2} are expected to
be valid only for $U\gg \Delta$, even for the moderate interaction of
Fig.~\ref{fig-fermicoeff}, surprisingly good agreement with the
complete solution is found for $|\varepsilon_d + U/2|\lesssim U/2$.  In
the opposite limit of an almost empty orbital, $\langle
\nd \rangle\approx 0$, interactions are negligible, and transport is well
described by a non-interacting resonant level model.  The crossover
from the local-moment to the empty-orbital regime becomes universal
for large values of $U$, for which the dimensionless Fermi-liquid
parameters, $\Delta \;\alpha_1$, $\Delta\; \phi_1$, $\Delta^2
\;\alpha_2$, and $\Delta^2 \;\phi_2$ can be expressed as
universal functions of $\varepsilon_d/\Delta$.

\begin{figure}[tbhp]
\includegraphics[width=0.9\columnwidth]{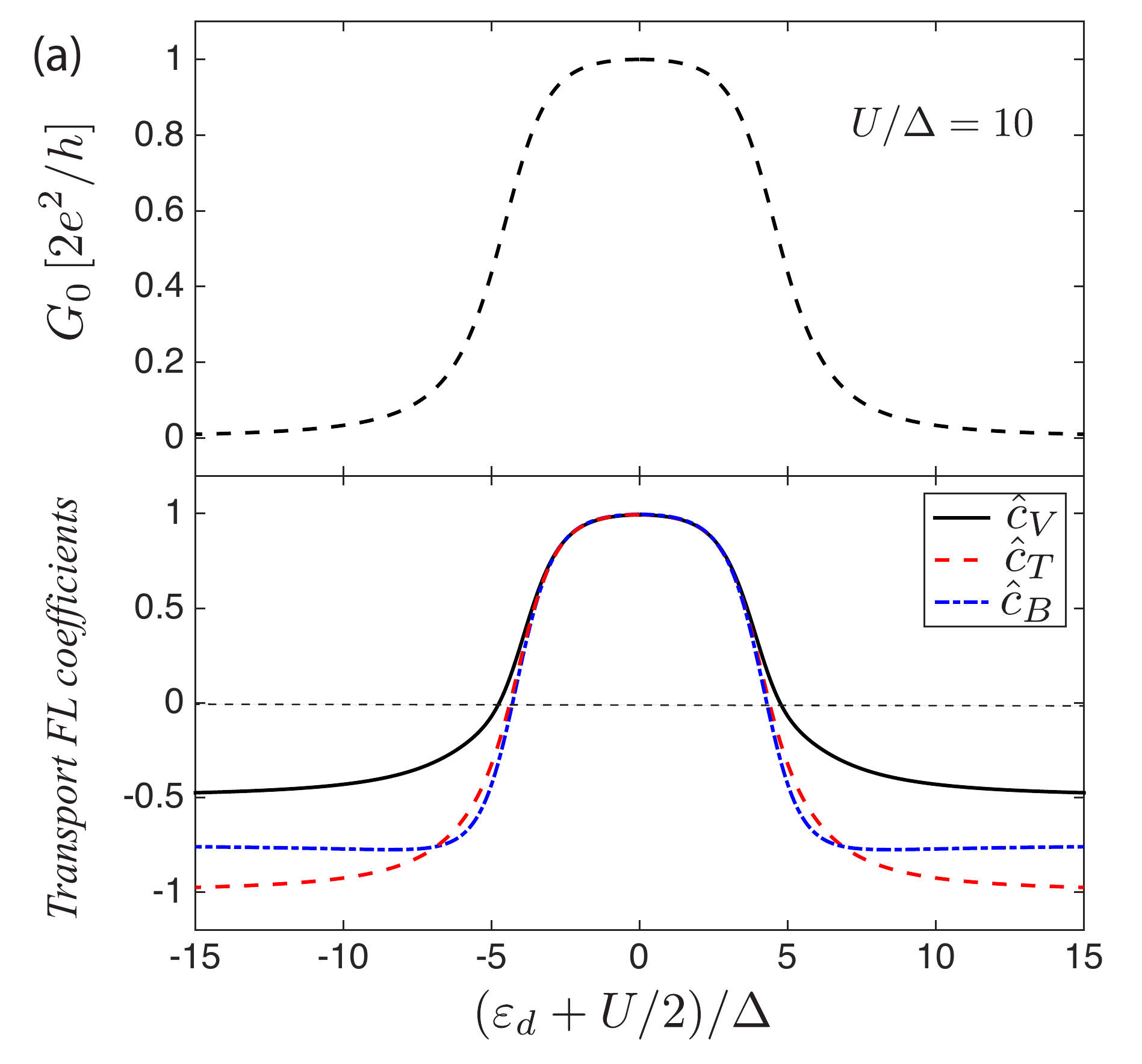}
\includegraphics[width=0.9\columnwidth]{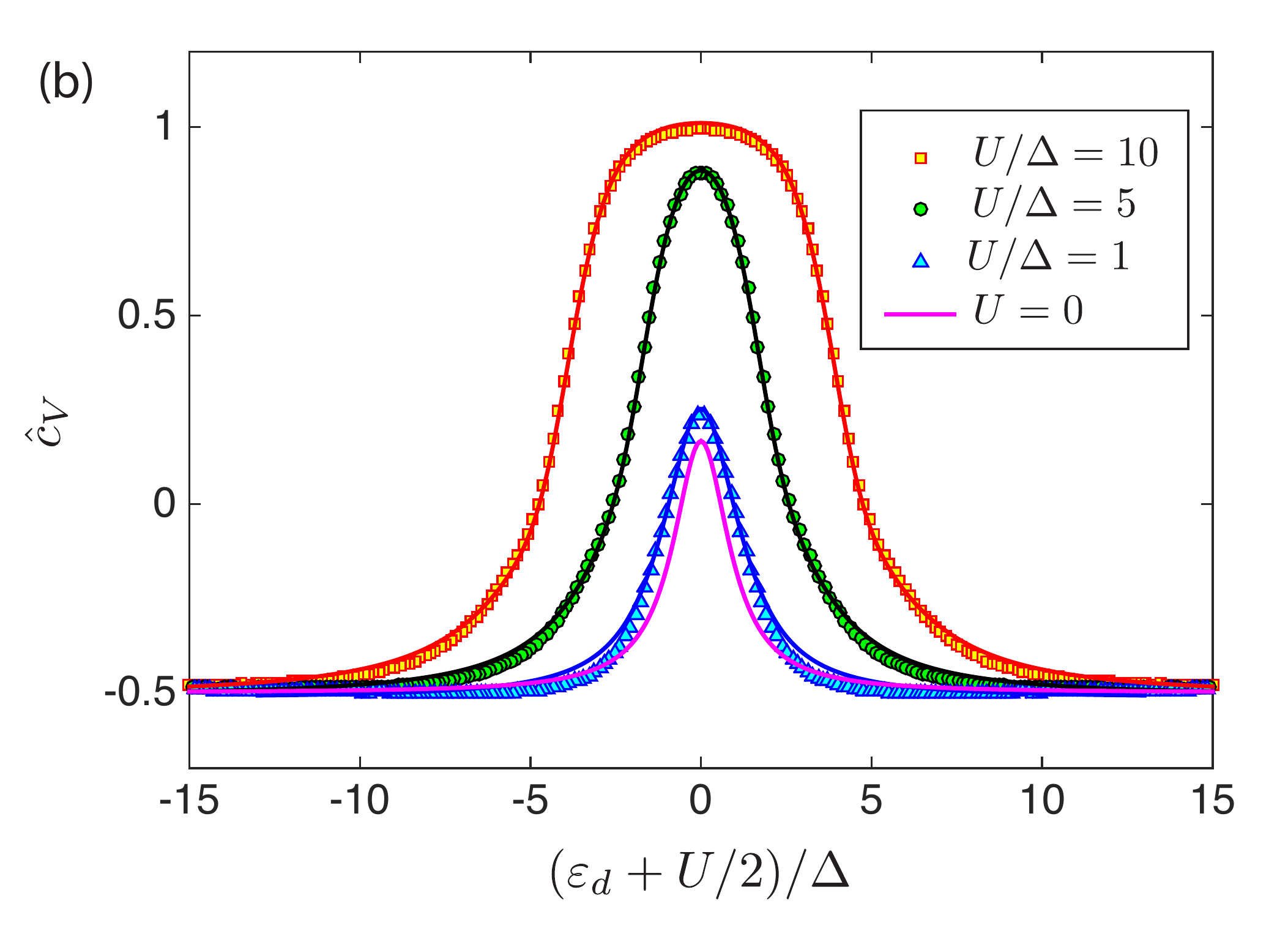}
\caption{\label{fig:cV_cT_cB}(Color online) (a) Normalized
  Fermi-liquid transport coefficients
  ${\hat c}_B \equiv c_B/c^{\rm K}_B $,
  ${\hat c}_T \equiv c_T/c^{\rm K}_T $, and
  ${\hat c}_V \equiv c_V/c^{\rm K}_V $ as a function of the level
  position for $U/\Delta = 10$,  \changes{obtained}
  from \changes{Bethe ansatz computations with} \Eqs{coeff-cb},
  (\ref{cT}) and (\ref{cV}).   The linear conductance $G_0$ is shown for
  comparison in the top panel, in units of $2 e^2/h$. \changes{(b)
    Transport Fermi-liquid coefficient $\hat{c}_V = c_V/c^{\rm K}_V$, plotted
    as function of $(\varepsilon_d+U/2)/\Delta$ for different values
    of $U/\Delta$, computed using the Bethe Ansatz (lines) and the numerical
    renormalization group (symbols).}  }
\end{figure}

Equipped with our Fermi-liquid theory and with the four Fermi-liquid
parameters, we then study a quantum dot device, coupled
symmetrically to two leads~\footnote{For asymmetrical lead coupling, 
we refer to Ref.~\cite{mora2009} where the calculations of the current and 
 noise have been carried out in great detail. Ref.~\cite{mora2009} 
uses the same FL Hamiltonian as in this work, albeit with different 
values of the FL parameters.}, and derive exact results for the FL
transport coefficients, $c_V$, $c_T$, and $c_B$, characterizing the
conductance at low bias voltage, temperature and magnetic field,
\begin{equation}
G (V,T,B)-G_0 \approx - \frac{2 e^2/h} {(\Estar)^2}
\left( c_T T^2 + c_V (e V)^2 
+c_B B^2\right),
\label{eq:c's}
\end{equation}
with $G_0= (2e^2/h)\sin^2(\delta_0)$ denoting the linear conductance
of the quantum dot at zero temperature and zero magnetic field.  In
terms of the Fermi-liquid parameters, the coefficient $c_B$ can be
expressed, e.g., as
\begin{equation}\label{coeff-cb}
\begin{split}
  c_B = \changes{-} \frac{\pi^2}{64 }\; \frac{\left( \alpha_2 + {\phi_2}/{4}
    \right) \sin 2 \delta_0 \changes{+} \left( \alpha_1 + \phi_1 \right)^2
    \cos 2 \delta_0}{\alpha_1^2}\;.
\end{split}
\end{equation}
The other two coefficients $c_V$ and $c_T$ are expressed by similarly
complex expressions, given by \Eqs{cT} and (\ref{cV}) in
Section~\ref{sec-FLcoeff}.  The value of these coefficients can be
trivially determined in the empty-orbital regime, where the following
asymptotic values are obtained, \be c^{\rm eo}_T = -\frac{\pi^4}{16},
\qquad c^{\rm eo}_V = - \frac{3 \pi^2}{64},\textrm{ \quad and\quad }
\changes{c^{\rm eo}_B = -\frac{3 \pi^2}{64}}\;.
\label{coefficients_in_empty_orbital_regime}
\ee Moving to the Kondo regime, the coefficients
$c_T$ and $c_V$ change sign and their ratio changes by a factor of 2
as compared to the empty-orbital regime, \be c^{\rm K}_T =
\frac{\pi^4}{16} \simeq 6.009, \quad c^{\rm K}_V = \frac{3 \pi^2}{32}
\simeq 0.925,
\label{eq:cv_ct_Kondo}
\ee
reflecting the emergence  of strong correlations in the Kondo   regime. In hindsight, this  sign change may be
not very surprising: in the Kondo regime, the perfect conductance through the Kondo resonance is reduced by a finite temperature (bias), destroying Kondo coherence, while in the empty-orbital regime a gradual lifting of the Coulomb blockade
 is expected as the temperature or bias voltage is increased.

\changes{$c_B$ also changes sign and its ratio with $c_V$ increases by a factor $3/2$ in the Kondo regime, where}
\begin{align}
 c^{\rm K}_B = \frac{\pi^2}{16}\simeq 0.617\; .
 \label{eq:cb_Kondo}
\end{align}
The evolution of the normalized coefficients $c_V/c_V^{\rm K}$,
$c_B/c_B^{\rm K}$, and $c_T/c_T^{\rm K}$ is shown in
Fig.~\ref{fig:cV_cT_cB}(a) for $U/\Delta=10$ as a function of the level's
position, $\varepsilon_d$, \changes{using Bethe ansatz computations. Susceptibilities
can also be computed from NRG and Fig.~\ref{fig:cV_cT_cB}(b) illustrates the
excellent agreement between Bethe ansatz and NRG on one transport coefficient.}
 Importantly, all three transport coefficients
can be, in principle, extracted from transport measurements, and thus
the predictions of this Fermi-liquid theory can be verified by
straightforward transport measurements~\cite{kretinin2011}.

\begin{figure}[tbhp]
\includegraphics[width=0.8\columnwidth]{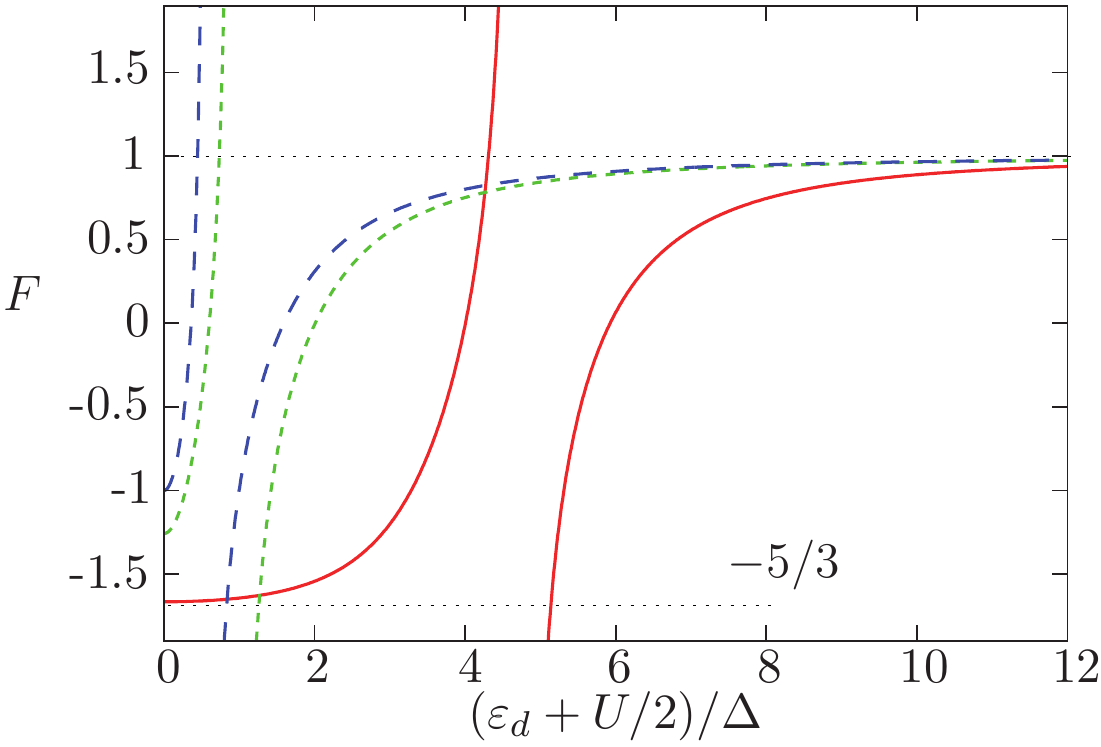}
\caption{\label{fig:fano} 
(Color online) Generalized Fano factor, 
Eq.~\eqref{fanofactor},  as functions of
  $(\varepsilon_d+U/2)/\Delta$ for $U/\Delta=10,1,0$ 
(full, dotted and dashed lines). The divergence of $F$ corresponds
to a vanishing current correction $\delta I$ , which occurs approximately
in the mixed-valence regime.}
\end{figure}

In addition, we also compute the zero frequency current noise at low voltage. 
It is characterized by a generalized Fano factor $F$~\cite{mora2009}, see Eq.~\eqref{fanofactor} in  Sec.~\ref{sec:fano}, defined as the ratio of the leading corrections to the noise and current 
with respect to the strong coupling fixed point values. 
We find for the Fano factor
\begin{equation}\label{fano}
F = \frac{\cos 4 \delta (\alpha_1^2+5 \phi_1^2) + 4 \phi_1^2 + \sin 4 \delta_0 
(\alpha_2/2-3 \phi_2/8)}{\cos 2 \delta_0 (\alpha_1^2+5 \phi_1^2) + \sin 2 \delta_0 
(\alpha_2-3 \phi_2/4)},
\end{equation}
displayed in Fig.~\ref{fig:fano} for different $U/\Delta$. At
particle-hole symmetry (in agreement with Ref.~\cite{sela2009}), $F$
varies between $-1$ in the non-interacting case $U=0$, corresponding
to Poissonian statistics for the backscattered current, to $-5/3$ at
large $U \gg \Delta$, emphasizing the role of interactions and
two-particle backscattering
processes~\cite{sela2006,gogolin2006,mora2009}. As $\varepsilon_d$
increases towards the empty orbital regime, the Fano factor
interpolates to the non-interacting Poissonian result $F=1$. The sign
change as $\varepsilon_d$ is varied indicates that $F$ describes a
backscattering current at $\varepsilon_d=-U/2$ but transmitted
electrons at large $\varepsilon_d$.

The rest of this paper is organized as follows.  In Sec.~\ref{sec:FL},
we construct the basic Fermi-liquid theory for the Anderson model and
relate the Fermi-liquid parameters of the effective Hamiltonian $\HFL$
to physical observables [\eqref{fermic}].  In Sec.~\ref{sec:current}
we construct the current operator and set the framework for
non-equilibrium calculations, which we then use to compute the
expectation value of the current and noise
perturbatively.  The final form of the transport coefficients
and Fano factor is presented in
Sec.~\ref{sec:FL_transport_coeff}. Sec.~\ref{sec:outlook}
  concludes and offers an outlook. The empty-orbital limit is discussed
in Appendix~\ref{appen-empty}. Technical details regarding the Bethe
  ansatz equations and their integral solutions, a Conformal Field
  Theory approach to the strong coupling fixed point and the
  calculation of the T-matrix, are left to the Supplemental
  Material~\cite{Note2}. In addition, the SM also contains detailed
  numerical results for the FL transport coefficients,
and a comparison to
  previous works for the Wilson ratio.

\section{Fermi-liquid theory}\label{sec:FL}

In this section, we present our Fermi-liquid theory for the Anderson
model. The Fermi-liquid theory is by essence a perturbative
approach. It gives the expansion of observables at bias voltages and
temperatures smaller than the Kondo temperature $T_K$. We begin in
Sec.~\ref{sec-noz} by a reminder of the Fermi-liquid approach to the
Kondo model, as introduced by
Nozi\`eres~\cite{nozieres1974,nozieres1978,nozieres1974b,lesage1999a,*lesage1999b},
and explain in detail how the model's invariance, in the wide-band
limit~\footnote{In the non-universal case of a finite
      bandwidth, the corrections to our predictions are expected to be
      small with the ratio of the maximum of $\Gamma$, $|\varepsilon_d|$
      and $U$, over the bandwidth of the model. }, 
under a global energy shift can be used to
relate the different Fermi-liquid parameters. In Sec.~\ref{sec-and},
we extend this approach to the Anderson model. In Sec.~\ref{sec-sus},
we take advantage of the Friedel sum rule to express all Fermi-liquid
parameters in terms of the spin and charge susceptibilities, see
Eqs.~\eqref{fermic}, a result of considerable practical
importance. The spin and charge susceptibilities are simple ground
state observables -- and can be computed semi-analytically by Bethe
Ansatz -- while the Fermi-liquid theory is able to deal with more
complicated situations, such as finite temperature or
out-of-equilibrium settings. Analytical expressions of the
Fermi-liquid parameters are obtained in the Kondo and empty-orbital
limits in Sec.~\ref{sec-anal}. Finally, the effective Fermi-liquid
Hamiltonian, applicable at low energy and already advertised in
Eq.~\eqref{eq:H_FL}, is discussed in Sec.~\ref{sec-hamil}.

\subsection{Kondo model}\label{sec-noz}

  We begin by briefly reviewing Nozi\`eres' local Fermi-liquid
  theory for the Kondo model. The main ideas are well established --
  for details we refer to the seminal papers of
  Nozi\`eres~\cite{nozieres1974,nozieres1978,nozieres1974b} or to
  Refs.~\cite{hewson1993,glazman2005,mora2009a}. Our goal here is to
  phrase the arguments in such a way that they will generalize
  naturally to the case of the Anderson model, discussed in the next
  subsection.

For energies well below the Kondo
temperature, the reduction of phase-space for inelastic processes
implies that elastic scattering dominates, due to the same phase-space
argument~\cite{landau1957a,*landau1957b,nozieres1964} as in
conventional bulk Fermi liquids. The system can then be characterized
by the phase shift, $\delta_\sigma (\varepsilon,n_{\sigma'})$,
acquired by a quasiparticle with kinetic energy $\varepsilon$ and spin
$\sigma$ that scatters off the screened Kondo singlet (the form of 
this phase shift can be derived
explicitly from the effective Fermi-liquid Hamiltonian Eq.~\eqref{eq:H_FL} [with
 $\alpha_2 = \phi_2 = 0$], as explained in Sec.~\ref{sec-hamil} below). Since the
singlet has a many-body origin, $\delta_\sigma
(\varepsilon,n_{\sigma'})$ depends not only on $\varepsilon$ but also
on the quasiparticle distribution functions $n_\uparrow
(\varepsilon')$ and $n_\downarrow (\varepsilon')$. Our goal is to find
a simple description of this phase shift function, valid for small 
excitation energies relative to the ground state.

In equilibrium and at zero temperature and magnetic field, the
quasi-particle ground state is characterized by a well-defined
zero-temperature chemical potential $\mu_0$.
Let $\varepsilon_0$ be an  arbitrary reference energy, different 
from $\mu_0$, which serves as the chemical potential of a reference ground state with
distribution function $n^0_{\varepsilon_0}(\varepsilon) = \theta
(\varepsilon_0 - \varepsilon)$. We then Taylor-expand the phase shift 
around this reference state as
\begin{equation}
\label{phshift}
\delta_\sigma (\varepsilon,n_{\sigma'}) = 
\delta_0 + \alpha_1 ( \varepsilon -\ezero)  -
 \phi_1 \int_{\varepsilon'} \delta n_{\bar{\sigma},\ezero} (\varepsilon') \; ,
\end{equation}
with $\delta n_{\sprime,\ezero} = n_{\sigma'} - \nzero_\ezero $. The
last term accounts for local interactions with other quasiparticles,
and $\bar{\sigma}$ denotes the spin opposite to $\sigma$, since by the
Pauli principle local interactions can involve only quasiparticles of
opposite spins. We should stress that the distributions $n_{\sigma'} 
(\varepsilon')$ \changes{can 
have arbitrary shapes (depending on chemical potential, 
temperature, magnetic field and, for out-of-equilibrium distributions,
source-drain voltage),} as long as the expansion variables
$\varepsilon-\varepsilon_0$ and $\int_{\varepsilon'} \delta n_{\bar{\sigma},\ezero} (\varepsilon')$
in Eq.~\eqref{phshift} are small compared to the Fermi-liquid 
scale $E^*$~\footnote{This implies in particular that the arbitrary 
energy $\varepsilon_0$ must be chosen near the zero-temperature chemical potential
$\mu_0$, $| \ezero - \muzero| \ll \Estar$.}. 
The Taylor coefficients $\delta_0$, $\alpha_1$ and
$\phi_1$ serve as the Fermi-liquid parameters of the theory. Their dependence
on $\varepsilon_0$ drops out in the wide-band limit considered here, and they are
universal coefficients.

Now, the key point is to realize that the function 
$\delta_\sigma (\varepsilon,n_{\sigma'})$ is of
course independent of the reference energy $\varepsilon_0$ used for its Taylor
expansion.
Differentiating Eq.~\eqref{phshift} w.r.t. $\varepsilon_0$ (and noting that
$\delta n_{\bar{\sigma},\ezero}  (\varepsilon')$ depends also on  $\varepsilon_0$)
 one thus obtains ${\rm d} \delta_\sigma (\varepsilon,n_{\sigma'})/{\rm d} \varepsilon_0 = \phi_1 - \alpha_1=0$, or
\begin{equation}
\label{FLiden}
\alpha_1 = \phi_1 \; .
\end{equation}
This relation constitutes one of Nozi\`eres' central
Fermi-liquid identities for the Kondo model. 

\begin{figure}[b]
\includegraphics[width=0.8\linewidth]{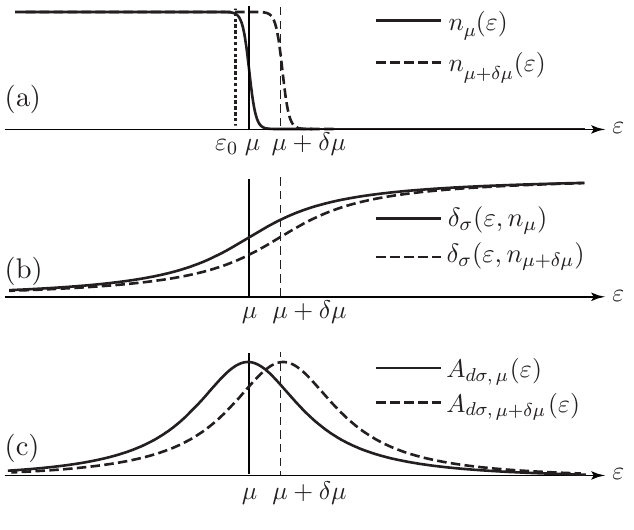}
\caption{\label{fig:shift-mu} Qualitative depiction of (a) the
  distribution function, (b) the phase shift and (c) the Kondo resonances in
  the impurity spectral function, for two choices of chemical potential,
  $\mu$ (solid lines) and $\mu + \delta \mu$ (dashed lines). Dotted
  lines illustrate the reference distribution function $n^0_\ezero
  (\varepsilon) = \theta (\ezero-\varepsilon)$ in (a).  }
\end{figure}

As can be checked easily, \Eq{FLiden} guarantees that for any
distribution $n_\sprime$ with a well-defined chemical potential, e.g.\
$n_\mu(\varepsilon') = (e^{(\varepsilon'-\mu)/T}+1)^{-1}$ for
nonzero temperature, the phase shift $\delta_\sigma(\varepsilon,
n_\mu)$, depends on energy and chemical potential only through the
combination $\varepsilon - \mu$. In other words, if $\mu$ is
changed to $\mu + \delta \mu$, e.g.\ by doping the system to
increase the electron density, then the new phase shift at
$\varepsilon + \delta \varepsilon$ equals the old one at
$\varepsilon$,
\begin{eqnarray}
\label{eq:floating-phase}
\delta_{\sigma} (\varepsilon + \delta \mu, n_{\mu + \delta \mu} ) = 
\delta_{\sigma}(\varepsilon, n_{\mu}) 
\; ,  
\end{eqnarray}
as illustrated in Fig.~\ref{fig:shift-mu}. [In fact, an alternative
way to derive \Eq{FLiden} is to impose \Eq{eq:floating-phase}, with
the same $\ezero$ on both sides of the equation, as condition on the
general phase shift expansion \Eq{phshift} for
$\delta_{\sigma}(\varepsilon, n_\mu)$; the calculations are simplest
if done at zero temperature, i.e.\ with $n_\mu \to \nzero_\muzero$.]
Since at $T=0$ the energy dependence of the phase shift determines
that of the Kondo resonance in the impurity spectral function, $A_{d
  \sigma, \mu} (\varepsilon)$, the latter, too, is invariant under a
simultaneous shift of $\varepsilon$ and $\mu$.  Pictorially speaking,
the ``Kondo resonance floats on the Fermi sea''
\cite{nozieres1974,mora2009a}: if the Fermi surface rises, the Kondo
resonance rises with it, and if the Fermi see is deep enough 
(wide-band limit), the Kondo resonance does not change its shape while 
rising.

The next step is to express $\delta_0$ and $\alpha_1=\phi_1$ in
terms of physical quantities, such as the local charge $n_d$ and the
local spin susceptibility $\chi_s$.  This can be done by calculating
the latter quantities via the Friedel sum rule,  evaluating the
ground state phase shift in a small magnetic field.  We discuss this
in detail in the next section, in the more general context of the
Anderson model.  Here we just quote the results: for the Kondo model,
one finds $\delta_0 = \pi/2$, $\alpha_1 = \phi_1 = \pi \chi_s$ and,
since $\chi_s=1/(4 \Tk)$, from Eq.~\eqref{eq:Delta*},
$\Estar = \Tk$ for the Fermi-liquid energy scale controlling the 
expansion Eq.~\eqref{phshift}.

Before proceeding further with the Anderson model, we wish to
emphasize two  important points: \\
(i) We have restricted our attention to elastic scattering
processes. As pointed out in Ref.~\cite{mora2009}, inelastic processes
involve the difference between the energies of incoming and outgoing
electrons and are therefore invariant under a global shift of all
energies by $\delta \mu$. \\
(ii) Eq.~\eqref{phshift} corresponds to the first few terms of a
general expansion of $\delta_\sigma (\varepsilon,n_{\sigma'})$ in
powers of $\varepsilon-\ezero$ and $\int_\eprime \delta
n_{\sigma',\ezero} (\varepsilon')$. In the calculation of the conductance,
for example at finite temperature, the $\alpha_1$ and $\phi_1$ terms
give a vanishing linear contribution and must therefore be taken into
account up to second order. To be consistent, one then needs to
include the next subleading terms $\sim 1/T_K^2$ in the expansion of
$\delta_\sigma (\varepsilon,n_{\sigma'})$. This has been worked out
explicitly for the SU($N$) case with
$N>3$~\cite{mora2008,mora2008b,vitushinsky2008,mora2009,mora2009a}. These
subleading terms, however, turn out to vanish identically in the SU(2)
Kondo model, as a result of electron-hole symmetry. This is no longer
the case for the asymmetric Anderson model, as we will see below.

\subsection{Anderson model}\label{sec-and}

 The Anderson model is described by a low-energy Fermi-liquid
  fixed point for all regimes of parameters, hence we now seek to
  generalize the above approach to this model, too.  The main
  complication compared to the Kondo model is that the Anderson model
  involves an additional energy scale, namely the impurity level
  $\varepsilon_d$, and its physics depends in an essential way on the
  distance $\ed-\muzero$ between its impurity energy level and the
  chemical potential.  
We again Taylor expand the phase
  shift w.r.t.\ to a reference energy $\ezero$, as in \Eq{phshift},
  but now include the next order in excitation energies
  \cite{mora2009a}: 
\begin{align}
\label{pshift-Anderson}
\delta_{\sigma} (\varepsilon,n_\sprime)
& = 
 \delta_{0,\ed-\ezero} +  \alpha_{1,\ed-\ezero} (\varepsilon-\varepsilon_0)  \\[1mm]
\nonumber
&  \phantom{.} \hspace{-10mm}
- \phi_{1,\ed-\ezero} \int_{\varepsilon'} \delta n_{\bar{\sigma},\ezero} 
(\varepsilon') +
\alpha_{2,\ed-\ezero}  (\varepsilon-\varepsilon_0)^2 \\[1mm] 
\nonumber
&  \phantom{.} \hspace{-10mm}
- \frac{1}{2} \phi_{2,\ed-\ezero}  
\int_{\varepsilon'} (\varepsilon + \varepsilon'- 2 \ezero) 
\delta n_{\bar{\sigma},\ezero} (\varepsilon') + \ldots 
\end{align}
$\delta_0$, $\alpha_1$, $\phi_1$, $\alpha_2$ and $\phi_2$ are the
Taylor coefficients of this expansion. In contrast to the
case of the Kondo model, they now \emph{do} depend explicitly on the
reference energy $\ezero$, and since we are in the wide-band limit,
this dependence can arise only via the difference $\ed - \ezero$.  For
notational simplicity, we will suppress this subscript below, taking
this dependence to be understood.  In the Kondo limit of
\Eq{eq:Kondo-limit}, the dependence on $\ed$ drops out, and the
coefficients $\delta_0$, $\alpha_1$, $\phi_1$, $\alpha_2$ and $\phi_2$
become universal, as seen in the previous section for $\delta_0$,
$\alpha_1$ and $\phi_1$.

Similarly to Sec.~\ref{sec-noz}, the Taylor coefficients are not all
independent as a result of the phase shift 
$\delta_{\sigma} (\varepsilon,n_\sprime)$ invariance under a change
in  $\varepsilon_0$.
 Differentiating Eq.~\eqref{pshift-Anderson} w.r.t. $\varepsilon_0$, and equating the coefficients of the various terms in the expansion (cst, $\sim (\varepsilon-\varepsilon_0)$, $\sim\int \delta n_{\bar{\sigma},\varepsilon_0})$ to zero, we therefore obtain the following three relations~\footnote{Further terms arise, proportional to
$\alpha'_2$ or $\phi'_2$ times term quadratic in $\varepsilon$, but we
ignore these, since they are of similar order as ones that would have
arisen had \Eq{pshift-Anderson} included terms cubic in $\varepsilon$,
which we neglected.}:
\begin{subequations}
\label{seteq}
\begin{eqnarray}
%
\label{seteq-a}
- \delta'_0 -  \alpha_1  + \phi_1 
& = & 0 \; , \\
\label{seteq-b}
- \alpha'_1 - 2 \alpha_2 + \phi_2 /2 
& = & 0 \; ,  \\[1mm] 
\label{seteq-c}
 \phi'_1+  \phi_2  & = & 0 \; .
\end{eqnarray}
\end{subequations}
Here a prime denotes a derivative with respect to the energy argument,
 e.g. $\delta'_0 = d(\delta_{0,\varepsilon_d - \varepsilon_0})/d
 \varepsilon_d$.

As can be checked easily, \Eqs{seteq} guarantee that for any
distribution $n_\sprime$ with a well-defined chemical potential, e.g.\
$n_\mu$, the phase shift $\delta_{\sigma, \ed} (\varepsilon,
n_\mu)$ (where the subscript $\ed$ indicates the
$\ed$ dependence of its Fermi-liquid parameters)
 remains invariant if $\varepsilon$, 
$\ed$ and $\mu$ are all shifted by the same amount:
\begin{eqnarray}
\label{eq:floating-phase-Anderson}
\delta_{\sigma,\ed+ \delta \mu } (\varepsilon + \delta \mu, n_{\mu + \delta \mu} ) = 
\delta_{\sigma,\ed}(\varepsilon, n_{\mu}) 
\; . 
\end{eqnarray}
Conversely, an alternative way to derive \Eqs{seteq} is to impose
\Eq{eq:floating-phase-Anderson} as a condition on the Taylor expansion
(\ref{pshift-Anderson}) for $\delta_{\sigma, \ed} (\varepsilon,n_\mu)$.

Collecting results, the first order
Fermi-liquid parameters, $\alpha_1$ and $\phi_1$, are related to each
other through
\begin{equation}
\phi_1 - \alpha_1 = \delta_0' \;  ,
\label{eq:phi1-alpha1}
\end{equation}
while the second-order Fermi-liquid parameters, $\alpha_2$ and
$\phi_2$, can be expressed via Eqs.~\eqref{seteq} in terms of
derivatives of lower-order ones:
\begin{equation}
\label{eq:phi2-alpha2}
\alpha_2 = -\frac{\delta_0^{\prime \prime}}{4}  - \frac{3 \alpha_1'}{4}, \qquad
\qquad \phi_2 = - \phi_1' . 
\end{equation}

Having established the above relations between the Fermi-liquid
parameters, we henceforth choose the reference energy at the
zero-temperature chemical potential, $\ezero = \muzero$.  Moreover,
since the choice of $\muzero$ is arbitrary in the wide-band 
limit, we henceforth set $\muzero=0$. Hence,
the energy argument of the Fermi-liquid parameters is henceforth understood 
to be $\ed$, i.e.\ $\delta_0$ stands for $\delta_{0,\ed}$, etc.

\subsection{Charge and spin static susceptibilities}\label{sec-sus}

\begin{figure}[b]
\includegraphics[width=\linewidth]{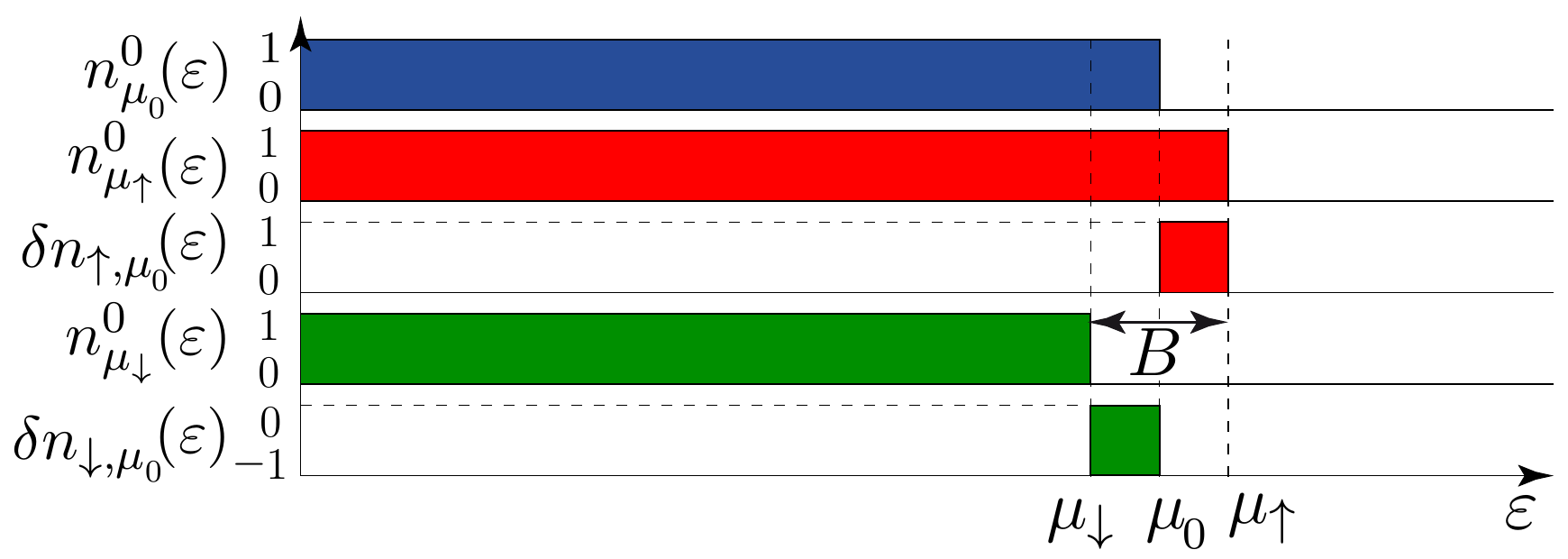}
\caption{\label{fig:spin} Zero-temperature quasiparticle
    distribution functions used for the calculation of
    Eq.~\eqref{eq:phshift-finite-B}: At zero field we use
    $\nzero_0$ as reference distribution (in Sec.~\ref{sec-sus},
    we set $\ezero = \muzero = 0$), while the system at small field $B$ has
    distribution $\nzero_{\mu_\sigma} $, differing from the reference
    distribution by $\delta n_{\sigma, 0} = \nzero_{\mu_\sigma}
    - \nzero_0$. The shifted chemical potentials, $\mu_\sigma =
    \sigma B/2$, derive from the condition $\langle
    b^\dagger_{\varepsilon \sigma} b_{\varepsilon \sigma} \rangle = 0$
    for $\varepsilon - \sigma B/2 > 0$.  }
\end{figure}

Our next task is to express the Fermi-liquid parameters in terms of
physical quantities. This can be done using the Friedel sum rule. To
this end, consider a zero-temperature system in a small nonzero
magnetic field, $B$, with distribution $\nzero_{\mu_\sprime}(\eprime)
= \theta(\mu_\sprime - \eprime)$ and spin-split chemical potentials,
$\mu_\sprime = \sigma' B/2$, as illustrated in Fig.~\ref{fig:spin}.
Using this distribution for $n_\sprime$ in \Eq{pshift-Anderson}, with
$\ezero = 0$ and $\delta n_{\bar \sigma, 0} = \nzero_{\mu_{\bar
    \sigma}} - \nzero_0$, we find:
\begin{eqnarray}
\label{eq:phshift-finite-B}
\delta_{\sigma} (\varepsilon,\nzero_{\mu_\sprime} ) & = & 
\delta_0 + \alpha_1 \varepsilon  -
\frac{\phi_1}{2} \bar \sigma  B + \alpha_2 \varepsilon^2 
\nonumber \\
& &   - \frac{\phi_2}{2}
\left[\varepsilon \bar \sigma B/2 + B^2 / 8 \right] 
 \; . \qqph
\end{eqnarray}
Now evoke the Friedel sum rule \cite{friedel1956}.  For given spin
$\sigma$ it relates the average charge bound by the impurity at $T=0$,
$\ndsigma = \langle \hatndsigma \rangle$, to the ground state phase
shift at the chemical potential, i.e.\ at $\varepsilon = \mu_\sigma$:
\begin{subequations}
\begin{eqnarray}
  \label{eq:Friedel}
\pi \ndsigma   &  = & 
\delta_\sigma (\mu_\sigma,\nzero_{\mu_\sprime} )
\\ 
\label{eq:Friedel-small-B}
& = & 
 \delta_0 + \frac{\sigma}{2} (\alpha_1 + \phi_1) B  
+ \frac{1}{4} ( \alpha_2 + \phi_2/4) B^2 
\; . \qqph
\end{eqnarray}
\end{subequations}
Thus, the average local charge $\nd$ and average magnetization $\md$
of the local level can be expressed as:
\begin{subequations}
\label{define-charge-magnetization}
\begin{eqnarray}
\nd & = & \ndup + \nddown = \frac{2 \delta_0}{ \pi} + 
\frac{1}{2 \pi} ( \alpha_2 + \phi_2/4) B^2 \; , \qquad
\\
\md & = & (\ndup - \nddown)/2 = \frac{ B}{2 \pi} (\alpha_1 + \phi_1) \; . 
\end{eqnarray}
\end{subequations}
In the strong-coupling Kondo regime we have $\nd
= 1$ at zero field, implying $\delta_0 = \pi/2$. 
In general, however, $\ndsigma$ is a function of $\ed$. From 
\Eqs{define-charge-magnetization}, the
local charge and spin susceptibilities at zero field
are given by 
\begin{subequations}
\label{eq:susceptibilities}
\begin{eqnarray}
\label{FLiden2}
\chi_c & = &  
- \left. \frac{\partial \nd}{\partial \ed}\right|_{B=0} 
= - 2 \frac{\delta'_0}{\pi} = 
\frac{2}{\pi} (\alpha_1 - \phi_1 ) \; , \\
  \label{spinsus}
  \chi_s & = & \phantom{-} \left. \frac{\partial \md}{\partial B} \right|_{B=0}  = 
 \frac{1}{2\pi} (\alpha_1 + \phi_1) \; . 
\end{eqnarray}
\end{subequations}
Using \Eqs{FLiden2} and (\ref{spinsus}), the  Fermi-liquid parameters can be
written in terms of the charge and spin susceptilibities $\chi_c$ and
$\chi_s$, and their derivatives w.r.t.\ to $\ed$, denoted by $\chi'_c
$ and $\chi'_s$.  The result is given in Eq.~\eqref{fermic} in the
introduction. As a consistency check, we note from 
Eq.~\eqref{fermic} that  $(\alpha_2 + \phi_2/4)/\pi = - \chi'_s$,
thus \Eqs{define-charge-magnetization} imply 
\begin{eqnarray}
  \label{eq:Friedel-consistency-check}
 \frac{\partial \nd }{\partial B} =  - 
 \frac{\partial \md }{\partial \ed} \; , 
\end{eqnarray}
which is a standard thermodynamic identity. 

For the Anderson model, $\nd$, $\chi_c$, $\chi_s$ and their
derivatives w.r.t.\ $\ed$ can all be computed using the Bethe Ansatz,
as detailed in the SM~\cite{Note2}.  This allows us to explicitly determine
how the Fermi-liquid parameters depend on $\ed$.  A corresponding plot
is shown in Fig.~\ref{fig-fermicoeff} for $U/\Delta=5$.

The Anderson model has a particle-hole symmetry,
which manifests itself as an invariance under the
  replacements $\varepsilon_d \to -\varepsilon_d -U$ for the impurity
  single-particle energy and $\nd \to 2-\nd$ for the impurity
  charge. The particle-hole symmetric point therefore corresponds to
  $\varepsilon_d = -U/2$ and $\nd=1$.  Moreover, $\chi_c$ and
    $\chi_s$ are symmetric with respect to particle-hole symmetry,
    while $\chi'_c$ and $\chi's$ are antisymmetric.  Consequently,
  Eqs.~\eqref{fermic} show that $\alpha_1$ and $\phi_1$ are symmetric
  while $\alpha_2$ and $\phi_2$ are antisymmetric, a feature already
  pointed out in the introduction. As a result, $\alpha_2$ and
  $\phi_2$ identically vanish at the particle-hole symmetric point
  $\varepsilon_d = -U/2$.  At this point, our result for the current
  will therefore agree with those of
  Refs.~\cite{nozieres1974,oguri2001,fujii2010,sakano2011}.  In the
  Kondo limit of \Eq{eq:Kondo-limit}, charge fluctuations are
  suppressed such that $\chi_c=0$, and Eq.~\eqref{FLiden2} reproduces
  the Fermi-liquid identity Eq.~\eqref{FLiden} of the Kondo model.

As discussed Section~S-1 of the Supplemental Material~\cite{Note2},
our approach reproduces the known FL relation between susceptibilities and the 
linear specific heat coefficient, and the corresponding Wilson ratio.

So far in this section, we have not used the specific form of the
Anderson model. The only ingredients that we have used are the
presence of a single-particle energy $\varepsilon_d$ for the impurity
and the assumption of Fermi-liquid behavior. This emphasizes the
generality of our Fermi-liquid approach, which is also applicable, for
instance, to other impurity models such as the interacting resonant
model~\cite{schlottmann1979}.

\subsection{Analytical expressions}\label{sec-anal}

  In order to better understand the
dependence of the Fermi-liquid parameters on $\varepsilon_d$, it is
instructive to consider certain limiting cases where analytical
expressions can be derived. In the Kondo regime, $U \gg \Delta$ and
$-U + \Delta < \varepsilon_d < -\Delta $, spin excitations
dominate and the charge susceptibility can be neglected ($\chi_c
  \simeq 0$, $\chi_c'\simeq 0$), so that 
[from \Eqs{fermic}]
\begin{eqnarray}
\label{eq:Fermi-liquid-parameters-Kondo-regime}
\alpha_1 \simeq \phi_1
  \simeq \pi \chi_s , \quad 4 \alpha_2/3 \simeq \phi_2 \simeq - \pi
    \chi'_s .
\end{eqnarray} 
The spin susceptibility is given
  with a very good accuracy by the asymptotical expression 
\begin{equation}
\label{asympto-kondo}
\chi_s = \frac{1}{2 \sqrt{2 U \Delta}} e^{\pi (U/8 \Delta -\Delta/2 U)} e^{-x^2},
\end{equation}
where we introduced the distance to the particle-hole symmetric point
$x = (\varepsilon_d + U/2) \sqrt{\pi/(2 \Delta U)}$. 
\Eq{asympto-kondo} agrees with the well-known formula $1/\Tk
  \propto (U \Delta)^{-1/2} e^{-\pi \ed (\ed + U)/(2 \Delta U)}$
  \cite{haldane1978}, up to an extra factor $e^{-\pi \Delta/(2U)}$,
  which was neglected in \cite{haldane1978} because the limit
  $U/\Delta \gg 1$ is implicit there.  Differentiating
Eq.~\eqref{asympto-kondo} w.r.t.\ $\varepsilon_d$, we find
\begin{equation}\label{asympto-kondo2}
   -  \chi'_s 
  = \frac{\pi^{1/2}}{2 \Delta U} e^{\pi (U/8 \Delta -\Delta/2 U)} x e^{-x^2}.
\end{equation}
\Eqs{eq:Fermi-liquid-parameters-Kondo-regime} to 
(\ref{asympto-kondo2}) together largely explain
the shape of all the curves in Fig.~\ref{fig-fermicoeff},
namely approximately Gaussian for $\alpha_1$ and $\phi_1$, 
or the derivative of a Gaussian for
$\alpha_2$ and $\phi_2$.

The other limit in which analytical expressions can be derived is the
empty-orbital regime, for $\varepsilon_d \gg \Delta$. The results are
detailed in Appendix~\ref{appen-empty}. Together with
Eqs.~\eqref{asympto-kondo} and~\eqref{asympto-kondo2}, they give us a
good analytical understanding of the $\varepsilon_d$ dependence
of the Fermi-liquid parameters. In the Kondo regime, $\alpha_1$ and
$\phi_1$ follow the spin susceptibility (or the inverse 
Kondo temperature) and decrease with increasing $\varepsilon_d$ (for
$\varepsilon_d > -U/2$) while crossing over into the
  mixed-valence regime. Finally, in the empty-orbital regime $\chi_s =
  \chi_c/4$, hence $\alpha_1$ still follows the spin susceptibility,
but with a factor $2$, $\alpha_1 \simeq 2 \pi \chi_s$, while $\phi_1$
becomes negligible. 

It is interesting to consider the ratios $\alpha_2/\alpha_1^2$ and
$\phi_2/\alpha_1^2$ which measure the importance of the second
generation of Fermi parameters compared to the first one. In the
Kondo region but far enough from particle-hole symmetry, $\alpha_2
\sim \phi_2 \sim 1/(T_K \Delta)$ [the precise formula is implied by
Eq.~\eqref{asympto-kondo2}] so that $\alpha_2/\alpha_1^2 \sim
\phi_2/\alpha_1^2 \sim T_K/\Delta$. The two ratios are small but
increase with $\varepsilon_d$ and $T_K$ towards the mixed-valence
region where they reach values of order $1$. Above, in the
empty-orbital region, $\varepsilon_d \gg \Delta$, $\phi_2/\alpha_1^2 =
6/\pi$ for $\varepsilon_d \ll U$ but is negligible for $\varepsilon_d
\gg U$, while $\alpha_2/\alpha_1^2 = \varepsilon_d/\Delta$ continues
to increase with $\varepsilon_d$ [see \Eqs{a12} to (\ref{b8})].

\subsection{Hamiltonian form}\label{sec-hamil}

The analysis carried out so far may seem abstract. It is based on the
elastic phase shift alone and it is not clear how transport quantities 
and other observables can be
computed. We thus need to write an explicit low-energy Hamiltonian
reproducing the phase shift of Eq.~\eqref{pshift-Anderson}.  The
leading order, or strong coupling Hamiltonian, is simply given by
the first term of Eq.~\eqref{eq:H_FL},
\begin{equation}\label{h0}
H_0 =  \sum_\sigma \int d \varepsilon \, (\varepsilon -\sigma B/2)  b_{\varepsilon \sigma}^\dagger b_{\varepsilon \sigma},
\end{equation}
where the quasiparticle operators $b_{\varepsilon \sigma}$, defined in the introduction, satisfy the fermionic anticommutation relations
\begin{equation}\label{anticom}
\{ b_{\varepsilon \sigma} , b_{\varepsilon' \sigma'}^\dagger \} = \delta_{\sigma,\sigma'} \delta (\varepsilon -\varepsilon'), \qquad \{ b_{\varepsilon \sigma} , b_{\varepsilon' \sigma'} \} = 0.
\end{equation}

The low-energy Hamiltonian admits an expansion in correspondence with
the phase shift expansion~\footnote{In Eq.~\eqref{eq:H_FL}, normal
  ordering is w.r.t. to a reference state with chemical potential
  $\varepsilon_0 = 0$. Its form for $\varepsilon_0 \neq 0$ is obtained
  by replacing $\varepsilon_i \to \varepsilon_i - \varepsilon_0$ in
  the coefficients of $\alpha_1$, $\alpha_2$ and $\phi_2$, and by
  normal ordering w.r.t. to $\varepsilon_0$.} of
Eq.~\eqref{pshift-Anderson}, the increasing orders being increasingly
irrelevant in the renormalization group
sense~\cite{lesage1999a,*lesage1999b}. The first two terms of this
expansion are given in Eq.~\eqref{eq:H_FL}. A more formal but
  complete justification of the form of the Hamiltonian, using
 conformal field theory arguments, is given in
  the SM~\cite{Note2}.

The computation of the elastic phase shift with $H$ involves all
processes stemming from $H_0$ and $H_\alpha$, in addition to the
Hartree diagrams inherited from $H_\phi$. Using $\delta_\sigma
  (\varepsilon)/\pi = \varepsilon - \sigma B/2 - \partial \langle \HFL
  \rangle / \partial n_\sigma (\varepsilon)$, it is straightforward to
check that Eq.~\eqref{pshift-Anderson} is reproduced, as required.

The low energy expansion of Eq.~\eqref{pshift-Anderson}
is valid as long as typical energies ($B$, $T$ or $V$) are smaller than 
a certain energy scale depending on $\varepsilon_d$. At large $U \gg \Delta$,
this energy scale is $T_K$ in the Kondo regime. It crosses over to $\Delta$
in the mixed-valence regime where physical quantities are universal when 
energies are measured in units of $\Delta$, see Sec.~S-II in~\cite{Note2}. 
In the empty-orbital regime, a resonant level model centered around $\varepsilon_d$
emerges, see appendix~\ref{appen-empty}, and this energy scale crosses over to
 $\varepsilon_d$.

To summarize this section, Eq.~\eqref{eq:H_FL} constitute a rigorous
and exact low-energy Hamiltonian for the Anderson model (or for other
similar models), and a basis for computing the low-energy quadratic
behavior of observables. We shall use it in the next section to
compute the conductance and the noise. The introduction of the elastic phase shift
was mainly aimed at determining the expressions of the Fermi-liquid
parameters given in Eq.~\eqref{fermic}.

\section{Current and noise calculations}\label{sec:current}

The Fermi liquid theory developed so far is very general, and applies
to many quantum impurity systems with a Fermi liquid ground state and
a single relevant channel of spinful electrons attached to it.  We now
turn to the concrete case of the Anderson model and calculate the
current and the noise through a quantum dot using the
Fermi-liquid theory described in the previous section.
For this purpose, the geometry of lead-dot coupling
  becomes important and scattering state wave functions have to be
  introduced in the spirit of Landauer's approach. Similar
calculations can be found in
Refs.~\cite{mora2008,mora2009,horig2014}. Sec.~\ref{sec-hamilt}
introduces the Anderson model and the corresponding Fermi-liquid
Hamiltonian valid at low energy, already outlined in the Introduction.
The current operator is given in Sec.~\ref{sec-current} and expanded
over the convenient basis of quasiparticle states. The perturbative
calculations of the current and noise current are then
separated into an elastic part in Sec.~\ref{sec-elastic} and an
inelastic part in Sec.~\ref{sec-inel}.

\subsection{Hamiltonians}\label{sec-hamilt}

\subsubsection{Anderson model}
\label{sec:Anderson-model-Hamiltonian}

We consider the model of a single-level dot symmetrically coupled to right and left leads with the Hamiltonian $H = H_{\rm a}+ H_{\rm AM}$, with $H_{\rm a} = \sum_\sigma \int d \varepsilon \, \varepsilon \, a_{\varepsilon \sigma}^\dagger a_{\varepsilon \sigma}$ and 
\begin{equation}\label{am}
\begin{split}
H_{\rm AM} &= \sum_\sigma \int d \varepsilon \, \varepsilon \,  \tilde{b}_{\varepsilon \sigma}^\dagger \tilde{b}_{\varepsilon \sigma}   +  \varepsilon_{d} \, \sum_\sigma n_{\sigma} \\ & + U \hatndup \hatnddown
+ \sqrt{\nu_0} \, t \sum_{\sigma} \int d \varepsilon \left( \tilde{b}_{\varepsilon \sigma}^\dagger  d_\sigma +  d_\sigma^\dagger
\tilde{b}_{\varepsilon \sigma}  \right),
\end{split}
\end{equation}
where, instead of the original left and right operators, $c_{{\rm
    L},\varepsilon \sigma}$ and $c_{{\rm R},\varepsilon \sigma}$, we
use the symmetric and antisymmetric combinations
\begin{equation}\label{rota}
\begin{pmatrix} \tilde{b}_{\varepsilon \sigma} \\ a_{\varepsilon \sigma} \end{pmatrix}
= \frac{1}{\sqrt{2}} \begin{pmatrix} 1 & 1 \\ 1 & - 1 
  \end{pmatrix}
\begin{pmatrix} c_{{\rm L},\varepsilon \sigma} \\ c_{{\rm R},\varepsilon \sigma} \end{pmatrix}.
\end{equation}
These satisfy the same anticommutation relations as in
Eq.~\eqref{anticom}. The leads are approximated, as
  usual\cite{wilson1975,tsvelick1983}, by a linear spectrum with a
  constant density of states $\nu_0$ per spin species,
  otherwise the results would not be
    universal. $d_\sigma$ is the electron operator of the dot and
  $n_\sigma = d^\dagger_\sigma d_\sigma$ the corresponding density for
  spin $\sigma$. $U>0$ denotes the charging energy, $\varepsilon_{\rm
    d}$ the single-particle energy on the dot and $t$ the tunneling
  matrix element from the dot to the symmetric combination of
  leads. The antisymmetric combination $a_{\varepsilon \sigma}$,
  associated with the wavefunction
\begin{equation}\label{antisym}
\psi^a_{k\sigma}(x) =
\left( e^{i ( k_F +k) x} - e^{-i ( k_F +k) x} \right)/\sqrt{2}  
\end{equation}
for all $x$, decouples from the dot variables. Here $x<0$ describes
the left lead and $x>0$ the right lead, energies and wavevectors are
related through $\varepsilon = \hbar v_F k$. For simplicity, the whole
system is assumed to be one-dimensional. Being odd in $x$, this
wavefunction vanishes at the origin and is therefore not affected by
the Anderson impurity. We define the hybridization $\Delta = \pi \nu_0
t^2$ for later use.

\subsubsection{Effective low-energy Hamiltonian}
\label{sec:eff-low-energy-Hamiltonian}

At low energy, screening takes place and the Anderson model 
flows to a Fermi-liquid fixed point for all values of
$\varepsilon_d$, $U$ and $\Delta$. The Hamiltonian describing the
low-energy physics of Eq.~\eqref{am} is then given by $H_a +\HFL$,
with the Fermi-liquid Hamiltonian $\HFL$ for the even channel given by
Eq.~\eqref{eq:H_FL}.

The difference between the original operators $\tilde{b}_{\varepsilon
  \sigma}$ associated with symmetric combinations of lead states
  and the corresponding quasiparticle operators $b_{\varepsilon
  \sigma}$ is the zero-energy phase shift $\delta_0$, i.e.\ the phase
shift that arises for $H_\alpha = H_\phi = 0$.  Hence $b_{\varepsilon
  \sigma}$ is associated with the scattering state
\begin{equation}\label{sym}
\psi^b_{k\sigma}(x) =
 \begin{cases}  
( e^{i (k_F+k) x} - {\cal S}_0
e^{-i (k_F+k) x} )/\sqrt{2}  \quad x<0,
\\ ( e^{-i (k_F+k) x} - {\cal S}_0 e^{i (k_F+k) x} )/\sqrt{2} \quad x>0,
 \end{cases}
\end{equation}
with the S-matrix ${\cal S}_0 = e^{2 i \delta_0}$. In
  contrast, for the antisymmetric combination of lead states described
  by $a_{\varepsilon \sigma}$-operators, which decouple from the dot
  variables, the corresponding S-matrix is trivially equal to 1, i.e.\
  the corresponding scattering phase is zero.

\subsection{Current operator}\label{sec-current}

In a one-dimensional geometry, the local current operator is given by
\begin{equation}\label{cur1}
\hat{I} (x) = \frac{e \hbar}{2 m i} \sum_\sigma \left( 
\psi_\sigma^\dagger (x) \partial_x  \psi_\sigma (x) 
- \partial_x \psi_\sigma^\dagger (x) \psi_\sigma (x) \right)
\end{equation}
where $m$ is the electron mass. Various expressions for the current
can be derived depending on which basis of states it is expanded
in. Here we choose a basis adapted to the low-energy model, namely we
expand over the zero-energy scattering states
\begin{equation}\label{expfield}
\begin{split}
\psi_\sigma (x) & 
= \int d \varepsilon \sqrt{\nu_0} 
\left[ \psi^a_{k\sigma}(x) \, a_{\varepsilon \sigma}   + 
\psi^b_{k\sigma}(x) \, b_{\varepsilon \sigma} \right] \; . 
\end{split}
\end{equation}
with $\nu_0=  1/hv_F$ the density of states of  incoming quasiparticles.

A voltage bias applied between the two leads, $\mu_L - \mu_R = e V$,
drives a current through the quantum dot. In a stationary situation,
the current is conserved along the one-dimensional space. We thus
define the symmetric current operator as $\hat{I} = (\hat{I} (x) +
\hat{I} (-x))/2$, where $x$ is arbitrary, corresponding to the average
of the left and right currents. Inserting the expansion
Eq.~\eqref{expfield} in Eq.~\eqref{cur1}, one finds the
Landauer-Buttiker~\cite{blanter2000} type current expression
\begin{equation}
\hat{I} = \frac{e}{2 h} \sum_\sigma \int_{\varepsilon,\varepsilon'} a_{\varepsilon \sigma}^\dagger  b_{\varepsilon' \sigma}  \left( e^{i(k'-k)x} -  {\cal S}_0 \, e^{-i(k'-k)x} \right) + {\rm h.c.},
\end{equation}
with $x<0$. A more compact expression can be obtained with the definition 
$a_{\sigma} (x)  \equiv \int d \varepsilon a_{\varepsilon \sigma} e^{i k x}$, namely
\begin{equation}\label{currentope}
\hat{I} = \frac{e}{2 h} \sum_\sigma  \left( a_{\sigma}^\dagger  (x) b_{\sigma}  (x) -   \,  a_{\sigma}^\dagger  (-x) \,({\cal S}_0b_{\sigma})(-x) + {\rm h.c.} \right). 
\end{equation}
Physically, operators taken at $x$ ($-x$) correspond here to incoming (outgoing) states.

Fluctuations in the current are characterized by the zero frequency current noise
\begin{equation}
S = 2 \int d t \langle \Delta \hat{I} (t) \Delta \hat{I} (0) \rangle
\end{equation}
where $\Delta \hat{I} (t) = \hat{I} (t) - \langle \hat{I} (t) \rangle$.

\subsection{Elastic scattering}\label{sec-elastic}

We study the average current through the dot in the presence of a voltage bias. We include in this section only the elastic and Hartree contributions, the inelastic terms will be considered in the next Sec.~\ref{sec-inel}. 

\subsubsection{Strong coupling fixed point}

We start by considering the strong coupling fixed point, {\it i.e.}
without the Fermi-liquid corrections $H_\alpha$ and $H_\phi$, where we
have a free gas of quasiparticles. The Hamiltonian is $H_0+H_{\rm a}$
and $a_{\varepsilon \sigma}^\dagger$ and $b_{\varepsilon'
  \sigma}^\dagger$ create eigenstates of the model. The left and right
scattering states, which are even and odd combinations of
$a_{\varepsilon \sigma}$ and $b_{\varepsilon' \sigma}$, are in thermal
equilibrium with spin-dependent chemical potentials
  $\mu_{L\sigma} = \mu_L + \sigma B/2$ and $\mu_{R\sigma} = \mu_R +
  \sigma B/2$.  Hence, we have
\begin{equation}
\label{eq:occupancies-aa-bb}
\begin{split}
  \langle a_{\varepsilon \sigma}^\dagger a_{\varepsilon' \sigma'}
  \rangle = & \langle b_{\varepsilon \sigma}^\dagger b_{\varepsilon'
    \sigma'} \rangle = \delta_{\sigma,\sigma'} \delta (\varepsilon
  -\varepsilon') \frac{f_{L\sigma} (\varepsilon) + f_{R \sigma}
    (\varepsilon)}{2} \\
  & \langle a_{\varepsilon \sigma}^\dagger b_{\varepsilon' \sigma'}
  \rangle = \delta_{\sigma,\sigma'} \delta (\varepsilon -\varepsilon')
  \frac{f_{L\sigma} (\varepsilon) - f_{R\sigma} (\varepsilon)}{2}
\end{split}
\end{equation}
with the Fermi distributions $f_{L\sigma} (\varepsilon)$ and
$f_{R\sigma}(\varepsilon)$. The mean value of the current $\hat{I}$
for the case of purely elastic scattering discussed in
  this subsection is then given by
\begin{equation}\label{cur2}
I = \langle \hat{I} \rangle = \frac{ e}{h} \sum_\sigma
  \int d \varepsilon {\cal T}_{\sigma} (\varepsilon) \bigl[ f_{L\sigma}
  (\varepsilon) - f_{R\sigma} (\varepsilon) \bigr] 
\end{equation}
with the transmission $ {\cal T}_\sigma (\varepsilon) = \sin^2(
\delta_0 )$, which here is energy- and spin-independent, because
$H_\alpha$ and $H_\phi$ have been neglected.  Performing the summation
over $\varepsilon$, one finds the average elastic
current 
$$
I = I_0 = (2 e^2 V/h) \sin^2 (  \delta_0)\;,
$$ 
which is maximal (unitary) at particle-hole symmetry
  $\delta_0 = \pi/2$ and approaches zero as $|\varepsilon_d -
  U/2|/\Delta$ becomes very large, so that $|\delta_0| \to 0$.

Correspondingly, the result for the noise is 
\begin{equation}\label{noise}
S = \frac{4 e^2}{h} \sum_\sigma \int_{\mu_{R\sigma}}^{\mu_{L\sigma}} 
d \varepsilon {\cal T}_\sigma (\varepsilon)
\left( 1 - {\cal T}_\sigma (\varepsilon) \right),
\end{equation}
and the partition noise is
$$
S = S_0 = (e^3 |V|/h) \sin^2 (2 \delta_0)
$$ 
at the strong coupling fixed point.

\subsubsection{Elastic scattering and phase shift}

We now include the Fermi-liquid terms $H_\alpha$ and $H_\phi$ into the
Hamiltonian. We first consider the elastic scattering processes
associated with $H_\alpha$.  Since they describe single-particle
processes, they can be absorbed in $H_0$ by a change of scattering
basis. The above analysis for computing the current and noise
can be reproduced
with the only change that the ${\cal S}$ matrix now carries an energy
and spin dependence, ${\cal S}_\sigma (\varepsilon) = e^{2
    i \delta_\sigma (\varepsilon)}$, and the knowledge of the phase
shift $\delta_\sigma (\varepsilon)$ suffices to characterize elastic
scattering. The resulting current and noise are still given by Eq.~\eqref{cur2}
and Eq.~\eqref{noise},
with $ {\cal T}_\sigma (\varepsilon) = \sin^2[\delta_\sigma
  (\varepsilon)]$.

Before writing the expression of the elastic phase shift, we note that
the Hartree terms stemming from $H_\phi$ are formally equivalent to
elastic scattering. Diagrammatically, each interaction vertex
connecting a fermionic line to a single closed fermionic loop (a
bubble) is similar to a local potential vertex where the energy is
conserved after scattering. As mentioned already
earlier, collecting purely elastic and Hartree
contributions, and calculating the phase shift, we indeed arrive at
Eq.~\eqref{pshift-Anderson}. 

For the rest of this section, we set $B=0$. At finite
temperature $T$ and voltage $V$, the
energy integrals in the phase shift expansion
Eq.~(\ref{pshift-Anderson}) yield
\begin{equation}
\int_\varepsilon \delta n_{\sigma,0} (\varepsilon) = 0 , 
\qquad \int_\varepsilon \varepsilon \delta n_{\sigma,0} (\varepsilon) 
= \frac{(\pi T)^2}{6} + \frac{(e V)^2}{8},
\end{equation}
so that we obtain the 
spin-independent phase shift
\begin{equation}\label{pshift3}
\delta_\sigma (\varepsilon)
=  \delta_{0}   + \alpha_{1} \varepsilon + \alpha_{2}  \varepsilon^2 
 -  \phi_2 \left( \frac{(\pi T)^2}{12} + \frac{(e V)^2}{16} \right).
\end{equation}
Inserting this result into Eq.~(\ref{cur2})
 for the elastic current and expanding to third order in energy, one obtains
\begin{equation}\label{iel}
\begin{split}
& I_{\rm el}  = \frac{2 e^2 V}{h} \Bigg[ \sin^2 \delta_0 - \sin 2 \delta_0 \, \phi_2  \left( \frac{(\pi T)^2}{12} + \frac{(e V)^2}{16} \right) \\[2mm]
& + \left( \alpha_2 \sin 2 \delta_0 + \alpha_1^2 \cos 2 \delta_0 \right) \left( \frac{(\pi T)^2}{3} + \frac{(e V)^2}{12} \right) \Bigg]\; . 
\end{split}
\end{equation}
This represents the elastic and Hartree contributions to the
current.

For the noise, we find $S = S_0 + \delta S_{\rm el}$ with
\begin{equation}\label{sel}
\frac{\delta S_{\rm el}}{4 e^5 |V|^3/h} = \frac{\alpha_1^2}{12} \cos 4 \delta_0 + \sin 4 \delta_0 \left(\frac{\alpha_2}{24}-\frac{\phi_2}{32} \right).
\end{equation}

\subsection{Inelastic scattering}\label{sec-inel}

\begin{figure}
\includegraphics[width=3.5cm]{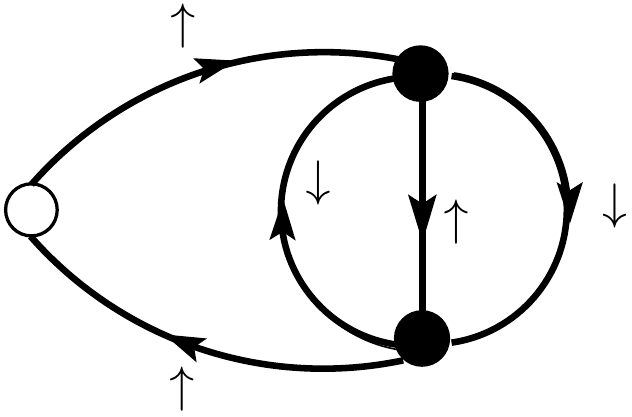}
\caption{\label{fig-diag} This diagram represents an inelastic process in which an electron is scattered and locally excites an electron-hole pair.}
\end{figure}

In the previous section, only the Hartree diagrams associated to
$H_\phi$ and the terms $H_\alpha$ have been included in the current calculation. A full account
of $H_\phi$ requires the use of the Keldysh
framework~\cite{kamenev2009} to compute the current in an
out-of-equilibrium setting. The average current is given by
\begin{equation}\label{current2}
I = \langle T_c \hat{I} (t) e^{-\frac{i}{\hbar} \int_{\cal C} d t' :H_\phi:(t')} 
\rangle,
\end{equation}
where $:H_\phi:$ denotes the interaction terms $H_\phi$ in Eq.~\eqref{eq:H_FL}, with the Hartree 
contributions removed and incorporated in the scattering wave functions and operators appearing in $H_0$.
The Keldysh contour ${\cal C}$ runs along the forward time
direction on the branch $\eta=+$ followed by a backward evolution on
the branch $\eta=-$, and $T_c$ is the corresponding time ordering
operator. Time evolution and mean values are determined by the free
Hamiltonian $H_0$, Eq.~\eqref{h0}, now incorporating  all elastic and Hartree
processes. Hence the current operator is given
by Eq.~\eqref{currentope} with $ {\cal S}_0$ simply replaced by the
energy-dependent $ {\cal S}_\sigma (\varepsilon)$. Starting with
Eq.~\eqref{current2}, we expand to second order in $:H_\phi:$,
 and compute the resulting integrals in
Keldysh space. The first order term vanishes by construction, and  the only remaining second-order
term is shown in Fig.~\ref{fig-diag}. The resulting  current contribution is~\cite{mora2009}
\begin{equation}\label{iinel}
I_{\rm inel} = \frac{2 e^2 V}{h} \phi_1^2 \cos 2 \delta_0  \left( \frac{2 (\pi T)^2}{3} + \frac{5 (e V)^2}{12} \right).
\end{equation}
Terms proportional to $\sim \phi_1 \phi_2$ and $\sim \phi_2^2$ are not
included here, since they involve higher powers of $T$ and/or
$eV$. The same is true for third or higher order terms in the
    expansion of $:\! H_\phi \! :$, which are proportional
  to $\sim \phi_1^3$ at least. As $\phi_1$ has the dimension of an
  inverse energy, the corresponding leading contributions to $I_{\rm
    inel}$ scale as $V T^3$ or $V^4$, and are hence neglected in our
  approach.  The total average current is obtained by summing the
elastic and inelastic terms, $I = I_{\rm el} + I_{\rm inel}$.

The inelastic contribution to the noise involves six diagrams. They are represented and calculated in detail in Refs.~\cite{mora2008,mora2009}. 
The result is $S = S_0 + \delta S$ with $\delta S = \delta S_{\rm el} + \delta S_{\rm inel}$ and
\begin{equation}\label{sinel}
\frac{\delta S_{\rm inel}}{4 e^5 |V|^3/h} = \phi_1^2 \left( \frac{1}{3} + \frac{5}{12} \cos 4 \delta_0 \right).
\end{equation}

\section{Fermi-liquid transport coefficients and Fano factor}\label{sec:FL_transport_coeff}

In this section, we discuss the results for the current obtained at low energy in terms of Fermi-liquid transport coefficients $c_B$, $c_T$ and $c_V$ introduced in Eq.~\eqref{eq:c's}.
We also compute the Fano factor related to low voltage noise.

\subsection{Finite magnetic field}\label{sec:c_B}

In principle, the set of Fermi-liquid parameters derived above is not
essential for the calculation of the linear conductance at zero
temperature and finite magnetic field. In this regime, the ground
state is still a Fermi liquid, even at large magnetic field.
Moreover, although a finite magnetic field separates the chemical
  potentials of the two spin orientations, $\mu_\sigma = \sigma B/2$,
  it does not create room for particle-hole excitations (a term of
  order $V^3$ at least is necessary for particle-hole excitations).
Thus, the linear conductance is given by \Eq{cur2}, which  reduces to
\begin{equation}\label{lin}
G = \frac{e^2}{h} \sum_\sigma \sin^2[\delta_\sigma (\varepsilon=\mu_\sigma)] \; .
\end{equation}
For $B=V=0$, this relates the phase $\delta_0$ to a physical
  observable, namely the linear conductance.  More generally, the phase shifts occurring
in \Eq{lin} are related via the Friedel sum rule, \Eq{eq:Friedel}, to the
spin-dependent populations, $\delta_\sigma (\mu_\sigma) = \pi
\ndsigma$. These are static observables that can be computed directly
from Bethe-Ansatz or NRG techniques, hence \Eq{lin} can be evaluated
without resorting to our Fermi-liquid expansion of the phase shift.

We may nevertheless use the latter to compute the low-field
expansion of the linear conductance, as given by Eq.~\eqref{eq:c's},
in order to compare $c_B$ with $c_T$ and $c_V$.
Substituting the small-field Fermi-liquid expansion 
\Eq{eq:Friedel-small-B} for $\delta_\sigma(\mu_\sigma)$
into Eq.~\eqref{lin} and expanding in $B$
 we obtain the
Fermi-liquid coefficient $c_B$ given in Eq.~\eqref{coeff-cb}. 
This Fermi-liquid expression
interpolates continuously between the empty-orbital
Eq.~\eqref{coefficients_in_empty_orbital_regime} and Kondo limits
Eq.~\eqref{eq:cb_Kondo}. 

\subsection{Finite temperature and non-linear conductance}\label{sec-FLcoeff}

Since the definition of the Fermi-liquid scale is somewhat arbitrary, there is no unambiguous way to define the Fermi-liquid transport
coefficients $c_T$ and $c_V$ in the general case. Here we use the
definition of Eq.~\eqref{eq:c's} with the Fermi-liquid scale $\Estar$ defined in
 Eq.~\eqref{eq:Delta*}, which recovers conventional results in the
particle-hole Kondo limit where $\Estar= \Tk$. 
The current obtained in the previous section then yields the Fermi-liquid
transport coefficients
\begin{equation}
\label{cT}
c_T = \frac{\pi^4}{16}  \frac{\left( \frac{\phi_2}{12} - \frac{\alpha_2}{3} \right) \sin 2 \delta_0 - \left( \frac{\alpha_1^2}{3} + \frac{2 \phi_1^2}{3} \right) \cos 2 \delta_0}{\alpha_1^2}, 
\end{equation}
and
\begin{equation}\label{cV}
c_V = \frac{\pi^2}{64}  \frac{\left( \frac{3 \phi_2}{4} - \alpha_2 \right) \sin 2 \delta_0   -   \left( \alpha_1^2+ 5 \phi_1^2 \right) \cos 2 \delta_0}{\alpha_1^2}.
\end{equation}
At particle-hole symmetry, these expressions simplify since $\alpha_2
= 0$, $\phi_2 = 0$ and $\delta_0 = \pi/2$. They can be written in
terms of the Wilson ratio, $R= 1+\phi_1/\alpha_1$ [from
Eq. (2) in the SM~\cite{Note2}], namely $c_T = (\pi^4/48) [ 1+2
(R-1)^2]$ and $c_V = (\pi^2/64) [1+5 (R-1)^2]$. Their ratio is thus
given by
\begin{equation}
\frac{c_V}{c_T} = \frac{3}{4 \pi^2} \frac{1+5 (R-1)^2}{1+ 2 (R-1)^2},
\end{equation}
in agreement with
Refs.~\cite{oguri2001,sela2009,rincon2009,*rincon2009b,*rincon2010};
it interpolates between $3/(2 \pi^2)$ in the Kondo limit $R \to 2$ and
$3/(4 \pi^2)$ in the non-interacting limit $R \to 1$. The values of
$c_T$ and $c_V$ in the Kondo regime are given in
Eqs.~\eqref{eq:cv_ct_Kondo}.  In the non-interacting limit, $U=0$,
i.e.\ for the resonant level model, the FL transport coefficients are
readily calculated. \changes{Their ratios are found to be independent of $\varepsilon_d$,
$c_V/c_T = 3/(4 \pi^2)$ and $c_T/c_B = 4 \pi^2$, 
with $c_V = (\pi^2/64) (\Delta^2 - 3 \varepsilon_d^2)/(\Delta^2 + \varepsilon_d^2)$, see Fig.~\ref{fig:cV_cT_cB}(b).}

\subsection{Fano factor}\label{sec:fano}

Following Refs.~\cite{mora2008,mora2009,vitushinsky2008}, we introduce
a generalized Fano factor 
\begin{equation}\label{fanofactor}
F = \frac{1}{2 e} \left . \frac{\delta S}{\delta I}\right |_{V\to 0}\;\; ,
\end{equation}
comparing the leading non-linear parts of the noise and current expansion,
$\delta S = S - S_0$ and $\delta I = I-I_0$. We note that, with the exception
of the two limits $\delta_0 \to \pi/2$ and $\delta_0 \to 0$, the low-voltage
current and noise are dominated by their strong coupling values $S_0$ and $I_0$.

Collecting the results of the current and noise corrections, 
Eqs.~\eqref{iel}, \eqref{sel}, \eqref{iinel}, and~\eqref{sinel}, we find 
the expression Eq.~\eqref{fano} advertised in the introduction.

\section{Conclusion and outlook} 
\label{sec:outlook}

The quasiparticle Fermi-liquid theory presented here provides a simple
and controlled framework to describe the leading behavior of the
Anderson model at low temperatures, voltages, and magnetic
fields. 
It should also be possible to obtain the results presented here with
other methods such as renormalized perturbation theory
(RPT)~\cite{hewson1993,*hewson1993b,*hewson1994}. It is, however, not
quite clear how the five parameters $\alpha_{1,2}$ and $\phi_{1,2}$
and the phase shift $\delta_0$, characterizing the generic
quasiparticle Fermi-liquid theory would appear in RPT. Just as the
underlying Anderson model, RPT has typically three parameters in its
usual form, $\tilde \varepsilon_d$, $\tilde U$, and $\tilde \Delta$.  It
is not absolutely clear if these three parameters are sufficient to
obtain the correct low temperature behavior, or if, similar to the
quasiparticle Fermi-liquid theory, additional parameters need be
introduced.  The parameters $\alpha_{1,2}$ could be incorporated,
e.g., via an energy dependent hybridization, $\Delta\to
\Delta(\varepsilon)$, but the implementation of the irrelevant operator
$\phi_2$ does not seem to be entirely straightforward.  Also,
extracting additional parameters of RPT directly from the finite size
NRG spectrum~\cite{krishna1980,*krishna1980a} may run into technical
difficulties.

As an outlook, let us put our results in a more general context.
  First, our expressions for $c_V$, $c_T$ and $c_B$ in terms of the
  Fermi-liquid parameters $\chi_c$, $\chi_s$, $\chi'_c$, $\chi'_s$ and
  $\delta_0$ are exact results relating transport coefficients
  to zero-temperature, equilibrium physical observables.  Our result
  for $c_V$ is, to the best of our knowledge, the first exact
  result for a nonequilibrium transport property of the Anderson model
  away from particle-hole symmetry.  This result constitutes a
  benchmark against which approximate analytical or numerical
  treatments of the nonequilibrium Anderson model
  \cite{Eckel2010,Pletyukhov2012} could be tested.

  Second, we emphasize that the conceptual framework laid out in the
  present paper is not tied to the specifics of the Anderson model.
  It could be applied to any other model whose low-energy fixed point
  is in the same universality class as that of the Anderson model.
  This is the case if the following conditions are met: (i) The model
  involves scattering of spinful electrons off a spatially confined
  region of charge; (ii) the model has SU(2) symmetry; (iii) the
  ground state is a spin singlet; and (iv) the scattering matrix
  involves only one nontrivial scattering phase (in the sense
  discussed in Section~\ref{sec:eff-low-energy-Hamiltonian}).  One
  example other than the Anderson model is the interacting resonant
  level model~\cite{schlottmann1979}, as already mentioned
  earlier. Another example would be a multi-level quantum dot
    model of the type studied in Ref.~\cite{Karrasch2007},
  with dot-lead coupling constructed such that only
  left-right-symmetric combinations of lead states couple to the dot
  while the anti-symmetric ones decouple, so that the S-matrix has only one
  non-trivial phase. For such a model, conditions (i-iv) are
  satisfied and the model's  low-energy fixed point is in the
  same universality class as the Anderson model. 
   Suppose one has access to a method that reliably
  captures the many-body correlations of such a model at zero
  temperature, but that is not able to treat nonzero temperature or
  nonequilibrium situations.  (An example of such a method would be
  the functional renormalization group in the Matsubara formulation,
  used in \cite{Karrasch2007}.)  Then low-$T$, low-$V$ predictions
  could be obtained via our Fermi-liquid approach by proceding as
  follows: First, one could use the zero-temperature, many-body method
  to calculate the local charge per spin species as function of gate
  voltage and magnetic field. Next, one could extract the Fermi-liquid
  parameters of the system via \Eqs{fermic} and
  \eqref{eq:phshift-finite-B} to \eqref{eq:susceptibilities}. Finally,
  our Fermi-liquid theory could be used for $T \neq 0$ or $V \neq 0$
  to calculate $c_T$ and $c_V$ as function of gate voltage, thus
  predicting the system's behavior at low temperature or low
  source-drain voltage.

Third, we remark that at $T=V=0$ the system is
a Fermi liquid for \emph{arbitrary} magnetic fields,
not only small ones. Hence, it is possible to generalize
the Fermi-liquid theory presented above to arbitrary $B\neq 0$,
and to calculate, for example, the Fermi-liquid transport
coefficients $c_T$ and $c_V$ as functions of $B$. This
analysis will be published separately.
  
Fourth, it would be very interesting to generalize our approach to
situations where both eigenphases of the scattering matrix are
nontrivial. The number of Fermi-liquid parameters would increase, but
it should still be possible to relate them all to
ground state values of physical observables.  A prime candidate for
which this would be useful would be a quantum point contact showing
the 0.7-anomaly \cite{Thomas1996,Micolich2011}.
It was recently shown experimentally that at low excitation energies
the 0.7-anomaly displays Fermi-liquid behavior \cite{Bauer2013} rather
similar to that of the Kondo effect. This experimental result suggests
that it should be possible to describe the low-energy behavior of the
0.7-anomaly using a Fermi-liquid theory \`a la Nozi\`eres. In
particular, it would be of great interest to calculate $c_B$, $c_T$
and $c_V$ as functions of the gate voltage controlling the width of
the quantum point contact, since these quantities were measured in
great detail experimentally \cite{Bauer2013}.  This could possibly be
done within the conceptual framework developed here, suitably
generalized to involve two nontrivial scattering phases and an
arbitrary magnetic field.   In this way,
Fermi-liquid theory could be used very instructively to elucidate the
low-energy behavior of the 0.7-anomaly.

\emph{Acknowledgement:} We thank F. Bauer, J. Heyder,
  M. Kiselev and D. Schuricht for insightful comments and lively discussions. 
\changes{We thank P. Rosenberger for correcting an important sign error in
one of our formulas.}
This work has been supported by the Hungarian research fund OTKA under
grant Nos.  K105149,  by the UEFISCDI grant DYMESYS (ANR
2011-IS04-001-01, Contract No. PN-II-ID-JRP-2011-1), and the DFG
  via SFB-TR12, De730/4-3 and the NIM Cluster of Excellence.

 \vspace{1cm}

\appendix

\section{Empty-orbital regime}\label{appen-empty}

In this Appendix, we examine the empty-orbital regime $\varepsilon_d
\gg \Delta$ using standard perturbation theory
(Rayleigh-Schr\"odinger). The unperturbed state is for $t=0$ (or
$\Delta = 0$), it corresponds to an empty impurity level with a filled
zero-temperature Fermi sea. Perturbation theory is carried out with
respect to the tunneling of electrons between the impurity and the
conduction sea. The unnormalized wavefunction of the ground state $|
\psi \rangle$ is computed to third order in $t$. The impurity
occupancy is then given by
\begin{equation}
  \nd = \frac{\langle \psi | \hatnd | \psi \rangle}{\langle \psi  | \psi \rangle}.
\end{equation}

For $U \gg \varepsilon_d$, we obtain the asymptotic expressions
\begin{equation}\label{chic1}
\chi_c = \frac{2 \Delta}{\pi \varepsilon_d^2}
\left[1 + \frac{2 \Delta}{\pi \varepsilon_d} 
 \left\{ - \frac{3}{2} + \ln \left( \frac{ \varepsilon_{d}}{U} \right) 
\right\} \right] ,
\end{equation}
for the charge susceptibility and
\begin{equation}\label{chic2}
\chi_s = \frac{\Delta}{2 \pi \varepsilon_d^2} 
\left[ 1  + \frac{2 \Delta}{\pi \varepsilon_d} 
 \left\{ \frac{1}{2} + \ln \left( \frac{ \varepsilon_{d}}{U} \right) \right\}
\right],
\end{equation}
for the spin susceptibility, in agreement with
Haldane~\cite{haldane1978a}. Eq.~\eqref{chic1} and Eq.~\eqref{chic2}
can also be derived from the mixed-valence results,
Eq.~(S-16) and Eq.~(S-17) in the SM~\cite{Note2}, in the limit
$\varepsilon_{dR} \gg \Delta$.

In the opposite case $U \ll \varepsilon_d$, the results are
\begin{equation}\label{nimp}
\nd = \frac{2 \Delta}{\pi \varepsilon_d}  \left[
1 -  \frac{ \Delta U }{\pi \varepsilon_d^2} \right], 
\end{equation}
and
\begin{equation}\label{chisasymp}
\chi_s = \frac{\Delta}{2 \pi \varepsilon_d^2}  
\left[ 1 -  \frac{\Delta U }{ \pi \varepsilon_d^2}\right]. 
\end{equation}
The Fermi-liquid parameters can be deduced from these expressions
using Eqs.~\eqref{fermic}. To leading order in $\Delta/\varepsilon_d$
the parameters $\alpha_1$ and $\alpha_2$ that describe
elastic scattering do not depend on the ratio of
$U/\varepsilon_d$. They are given by
\begin{equation}\label{a12}
\begin{split}
\alpha_1 = \pi \left( \chi_s +  \frac{\chi_c}{4} \right) \simeq \frac{\Delta}{\varepsilon_d^2} , 
\\[2mm] \alpha_2  = - \pi \left( \frac{3}{4} \chi_s' + \frac{\chi_c'}{16} \right) \simeq \frac{\Delta}{\varepsilon_d^3},
\end{split}
\end{equation}
corresponding to the phase shift expansion of a non-interacting
resonant level model $\delta(\varepsilon) = {\rm atan}
[\Delta/(\varepsilon_d -\varepsilon)]$. The parameters $\phi_1$
  and $\phi_2$ that describe interaction processes depend on
$U/\varepsilon_d$. They are given by
\begin{equation}
\phi_1 = \pi \left( \chi_s - \frac{\chi_c}{4} \right) \simeq \frac{2 \Delta^2}{\pi \varepsilon_d^3} \quad \quad \phi_2 = - \phi_1' = \frac{6 \Delta^2}{\pi \varepsilon_d^4},
\end{equation}
for $U \gg \varepsilon_d$ and
\begin{equation}
\label{b8}
\phi_1 = \frac{\Delta^2 U}{\pi \varepsilon_d^4} \quad \quad \phi_2 = - \phi_1' = \frac{4 \Delta^2 U}{\pi \varepsilon_d^5},
\end{equation}
for $U \ll \varepsilon_d$. The corresponding FL transport coefficients
are given by Eq.~\eqref{coefficients_in_empty_orbital_regime}.


%

\clearpage

\section*{Fermi-liquid theory for the single-impurity Anderson model (Supplemental Material)}
\renewcommand{\thesection}{S-\Roman{section}}
\renewcommand{\theequation}{S-\arabic{equation}}
\renewcommand{\thefigure}{S-\arabic{figure}}

\setcounter{equation}{0}

Unless preceded by $S-$, cited equations refer to the main text.

\section{Wilson ratio}

 Let us here establish contact with previous works for the FL parameters $\alpha_1$
   and $\phi_1$.  The Friedel
    sum rule implies an impurity-induced change in density of states
    per spin species given by $\nu_{\sigma,\imp} =
    (1/\pi) \partial_\varepsilon
    \delta_{\sigma}(\varepsilon,\nzero_\muzero)|_{\varepsilon=\muzero}
    = \alpha_1/\pi$, and hence a corresponding impurity-induced change
    in the specific heat of $\gamma_\imp = (\pi^2 k_B^2/3) \sum_\sigma
    \nu_{\sigma , \imp} = (2 \pi k^2_B/3) \alpha_1$, where $k_B$
    denotes the Boltzmann constant.  Eliminating $\phi_1$ from
    \Eqs{eq:susceptibilities}, we find
\begin{eqnarray}
  \label{eq:4chis+chic}
  \frac{4 \chi_s}{(g \mu_B)^2} + \chi_c = \frac{4 \alpha_1}{\pi} = 
\frac{6 \gamma_\imp}{\pi^2 k_B^2} \; ,
\end{eqnarray}
where physical units have been reinstated (only in this equation) 
by replacing $\chi_s$ by 
$\chi_s/(g \mu_B)^2$. This relation agrees with previous Fermi-liquid
studies~\cite{yamada1975a,yosida1975,hewson1993,hewson1997kondo}. Next,
consider the Wilson ratio $R$, defined as the ratio of the impurity
contributions to the spin susceptibility and specific heat, $\chi_s$
and $\gamma_\imp$, relative to their respective bulk contributions,
$\chi_{s, \bulk} = \nu_0/2$ and $\gamma_\bulk = (\pi^2 k_B^2/3) 2
\nu_0$, where $\nu_0$ is the bulk density of states per spin
species. \Eq{eq:4chis+chic} implies
\begin{eqnarray}
  \label{eq:Wilson-ratio-definition}
R \equiv   \frac{ \chi_s/\chi_{s,\bulk}}{
    \gamma_\imp/\gamma_\bulk}   
=  
\frac{2}{1 + \chi_c/ (4 \chi_s)} \; , 
\end{eqnarray}
in agreement with Ref.~\cite{hewson1993}.  This interpolates between
the non-interacting case, where the charge and spin 
susceptibilities are trivially related by $\chi_s =  \chi_c/4$,
hence $R=1$, and the Kondo limit, where $\chi_c = 0$, hence  $R=2$.

\section{Numerical results for the FL transport coefficients}
\label{sec:results}

In the main text, we developed a quasiparticle Fermi-liquid
theory of the Anderson model. In its generic form, this Fermi-liquid
theory necessarily includes four Fermi-liquid parameters in addition
to the phase shift.  We used this Fermi-liquid theory to compute the
conductance through a symmetrically coupled quantum dot, and
determined the Fermi-liquid transport coefficients, $c_V$, $c_T$, and
$c_B$, defined in Eq.~\eqref{eq:c's}.  As we have shown in
Section~\ref{sec-sus} (already summarized in Eqs.~\eqref{fermic} of
the Introduction in the main text), the only inputs needed to compute the Fermi-liquid
coefficients, -- and thus the transport coefficients from
Eqs.~\eqref{coeff-cb}, \eqref{cT} and~\eqref{cV}, -- are the spin
($\chi_s$) and charge ($\chi_c$) susceptibilities and their
derivatives.  We obtained these susceptibilities via two complementary
methods: the Bethe Ansatz solution, discussed in
Sec.~\ref{append:ba}, and NRG~\cite{BudapestNRG}.  Extracting the
Fermi-liquid parameters from $\chi_s$ and $\chi_c$, we were then able
to compute the transport coefficients in terms of the bare parameters
of the Anderson model.

\changes{Our NRG computations were performed with a discretization
  parameter $\Lambda = 2$, while keeping 1024 states in each
  iteration.  In our computations we exploited the $U(1)\times U(1)$
  symmetry of the Hamiltonian, corresponding to the conservation of
  the charge $Q$ and the $z$-component of the spin $S_z$.  We used a
  flat band with half-width $D$ and density of states per spin species
  of $\nu_0 = 1/(2D)$, and fixed $\Delta=0.005 D$. 
   The charge
  susceptibility was computed simply as in Eq.~\eqref{FLiden2}, by
  taking the numerical derivative of the occupation $n_d$ with respect
  to $\varepsilon_d$. The spin susceptibility has been determined by
  applying a tiny magnetic field $B_z\sim 10^{-12}D \ll T_K$ and then
  making use of Eq.~\eqref{spinsus}.}
 
\begin{figure}[hptb]
\includegraphics[width=\columnwidth]{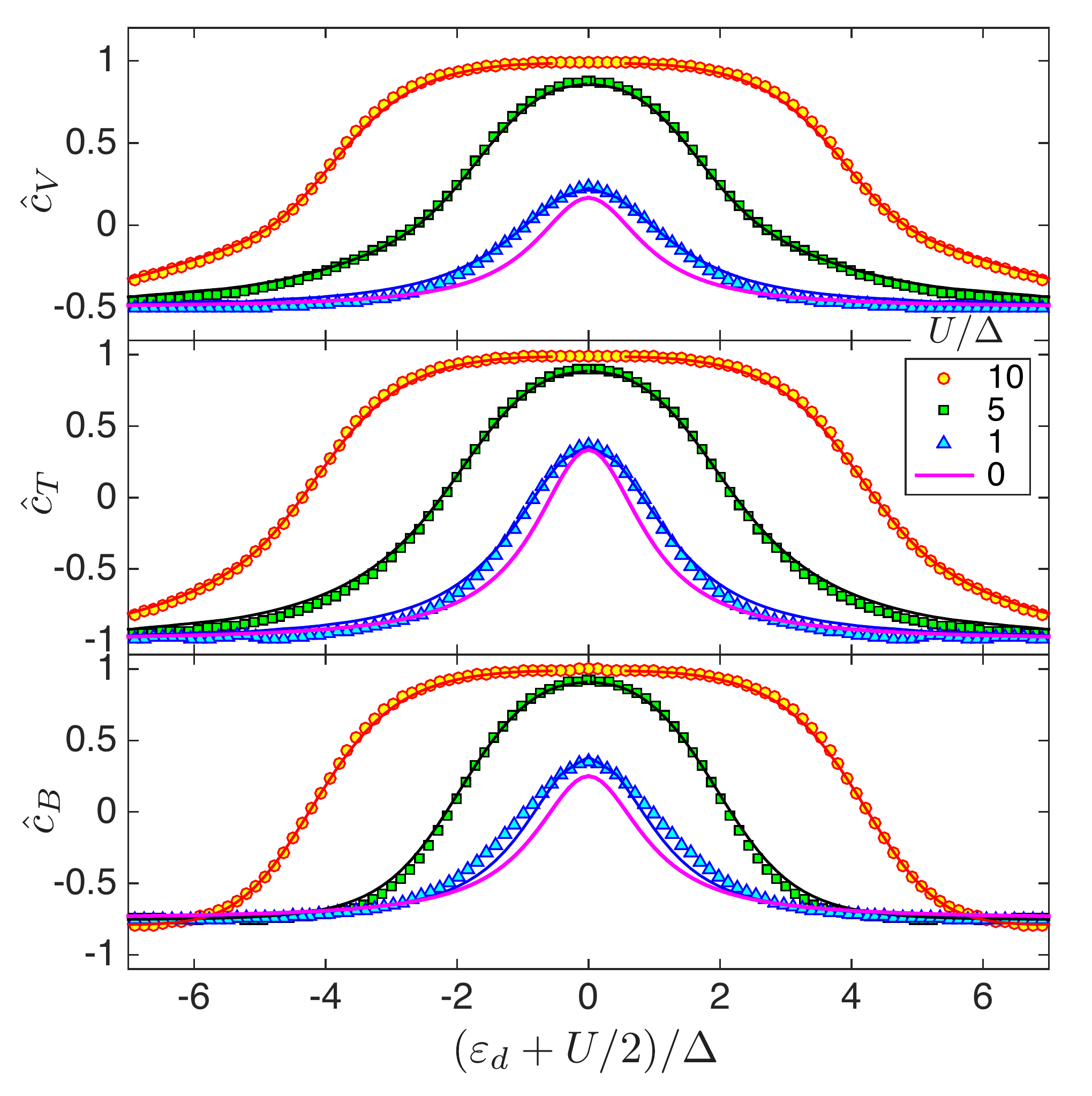}
\caption{\label{fig-cV} (Color online) The transport
  Fermi-liquid coefficients $\hat{c}_V = c_V/c_V^K$, $\hat{c}_T = c_T/c_T^K$ 
  and $c_B/c_B^K $, \changes{plotted as a
  functions of $(\varepsilon_d+U/2)/\Delta$ for different values of
  $U/\Delta$, computed using Bethe Ansatz (lines) and the numerical renormalization
group (symbols).}}
\end{figure}

\begin{figure}[tbhp]
\includegraphics[width=\columnwidth]{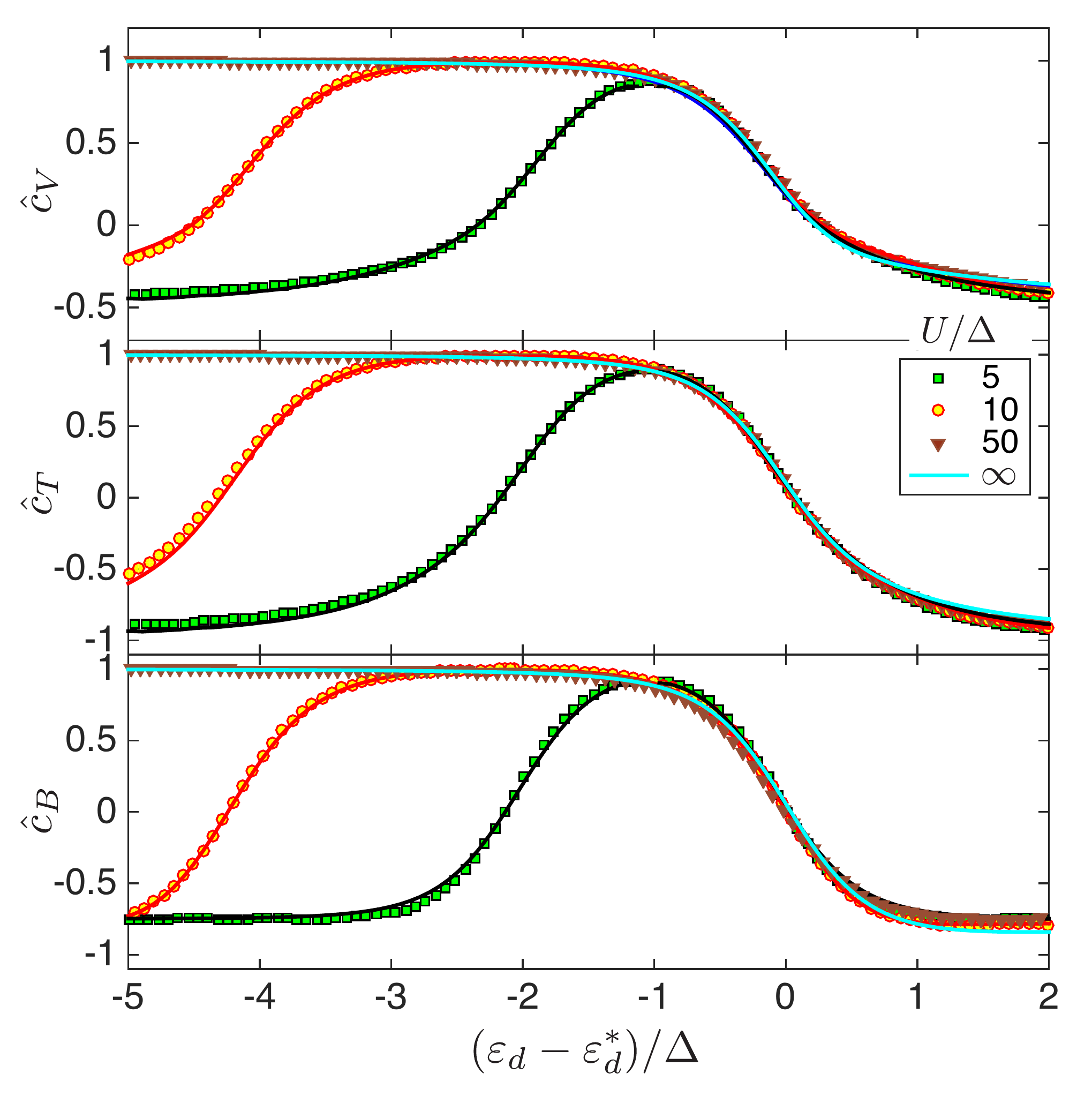}
\caption{\label{fig-scaling-cV}(Color online) Approach to the mixed 
valence regime $U\to \infty$ for the transport 
Fermi-liquid coefficients 
$\hat c_V$, $\hat c_T$, and $\hat c_B$,
computed using Bethe Ansatz (lines) and the numerical renormalization
group (symbols). The \changes{cyan} lines show the universal scaling
curves in the $U\to\infty$ limit.  By definition, the impurity
occupancy is $\nd =1/2$ for $\varepsilon_d = \varepsilon_d^*$.}
\end{figure}

The results for $c_B$, $c_T$ and $c_V$ were already advertised and
plotted in Fig.~\ref{fig:cV_cT_cB} in the main text for
$U/\Delta=10$.  The dependence of $c_V$ on the ratio $U/\Delta$ is
presented in Fig.~\ref{fig-cV} (reproducing Fig.~\ref{fig:cV_cT_cB}(b)). 
For $U/\Delta \lesssim 2$, the
$\varepsilon_d$ dependence of the coefficient $c_V$ is almost the same
as predicted by a non-interacting resonant level model. Notice that even in
this simple limit, $c_V$ does depend on the position of the resonant
level, since the slope and the curvature of the local density of
states both vary with the position of the level, $\varepsilon_d$.
Increasing the ratio $U/\Delta$ further, a local-moment regime
develops around $\varepsilon_d + U/2\approx 0$ for $U/\Delta \gtrsim
10$, where the value of the transport coefficients is approximately
given by Eqs.~\eqref{eq:cv_ct_Kondo} and \eqref{eq:cb_Kondo}.  The
size of the Kondo region (plateau) increases with $U/\Delta$, while
the crossovers from the Kondo to the empty-orbital regimes occur over
the energy scale $\Delta$.
As mentioned already in the Introduction in the main text, the crossover from the Kondo
regime to the empty-orbital regime becomes universal in the $U\to
\infty$ limit. To demonstrate this, we define the energy
$\varepsilon_d^*$ as the single-particle energy for which the impurity
occupancy is $\langle \hatnd\rangle =1/2$, and reproduce
Fig.~\ref{fig-cV} in Fig.~\ref{fig-scaling-cV}, but with the
single-particle energy $\varepsilon_d$ measured relative to
$\varepsilon_d^*$, and normalized by $\Delta$. Clearly, $c_V$ rapidly
approaches a universal crossover curve, $c_V =
f_V(\frac{\varepsilon_d-\varepsilon_d^*} \Delta)$ as the interaction
is increased. The scaling limit $U\to \infty$ can be accessed directly
in the Bethe Ansatz solution. In this case, the susceptibilities
$\chi_c$ and $\chi_s$ have integral representations (see
Eqs.~\eqref{mixed-chic} and \eqref{mixed-chis} in
Sec.~\ref{append:ba}), which can be used to compute the scaling
curves shown as continuous black lines in
Fig.~\ref{fig-scaling-cV}. The transport coefficients $c_T$ and $c_B$
exhibit similar scaling properties, shown in the lower two panels of
Figs.~\ref{fig-scaling-cV}.

The transport coefficients $c_V$, $c_T$, and $c_B$ are of immediate
experimental significance. Nevertheless, extracting their \emph
{absolute value} in a quantum dot experiment is not very
straightforward since, to do that, one should first determine the
scale $\Estar$ in Eq.~\eqref{eq:c's}, expressed from \eqref{fermic}
as 
\be 
\Estar = \frac{1}{4 \chi_s + \chi_c}\;.  
\label{eq:finalEstar}
\ee 
While measuring the gate voltage dependence of the charge on a quantum
dot and thus $\chi_c$ is not very difficult, it is extremely hard to
access the spin susceptibility $\chi_s$ in an ordinary quantum
dot. Both $\chi_c$ and $\chi_s$ can, however, be measured in a
spin-polarized capacitively coupled double quantum dot
device~\cite{amasha2013}, where charge degrees of freedom play the
role of ordinary spin.  In a large magnetic field, only spin-up electrons
can stay on each quantum dot, and the number of electrons on the left
and right dots, $\hat n_L = d^\dagger_{L\uparrow}
d^{\phantom{\dagger}}_{L\uparrow}$ and $\hat n_R =
d^\dagger_{R\uparrow} d^{\phantom{\dagger}}_{R\uparrow}$ play the same
role as $\hatndup = d^\dagger_{\uparrow}
d^{\phantom{\dagger}}_{\uparrow}$ and $\hatnddown =
d^\dagger_{\downarrow} d^{\phantom{\dagger}}_{\downarrow}$ in the
Anderson model.  In this case, both $\chi_c$ and $\chi_s$ can be
determined from the side gate dependence of the occupations $\langle
\hat n_R\rangle $ and $\langle \hat n_L\rangle $, monitored e.g. by
point contact sensors.
%


\section{Bethe Ansatz solution}\label{append:ba}

\subsection{Linear system}

An exact solution to the ground state of the Anderson model can be
derived using the Bethe Ansatz~\cite{tsvelick1983}. The description
involves spin excitations with wavevector $\lambda$, corresponding to
bound state singlet pairs, and unbound charge excitations with
wavevector $k$. The densities of states $\sigma(\lambda)$ and $\rho
(k)$ of these two types of excitations satisfy linear integral
equations (to be written below) that can be solved either numerically
or analytically in some parameter region with the help of the
Wiener-Hopf method~\cite{tsvelick1983}. The system described by the
spin and charge densities $\sigma(\lambda)$ and $\rho (k)$ corresponds
to $N$ electrons occupying either the dot single-level or the
one-electron states of the conduction band. Since we consider a large
number of electrons $N \gg 1$, the presence of the dot gives a
subleading contribution to the densities
\[
\sigma (\lambda)= \sigma_c (\lambda) + \frac 1 L 
\sigma_i (\lambda), \qquad 
\rho (k)= \rho_c (k) + \frac 1 L  
\rho_i (k),
\]
where the subscript $c/i$ stands for conduction/impurity (dot), $L$ is the system size increasing linearly with $N$.

$\sigma_c (\lambda)$ and $\rho_c (k)$ are the spin and charge
densities in the absence of the dot. They describe, in fact, a free
electron gas but in a complicated way.
 They are related to the external magnetic field $B$ and the parameters of the Anderson model~\cite{kawakami1982ground}
\[
\frac{B}{2 \pi} = \int_{-\infty}^\Lambda \rho_c (k) d k, 
\qquad \frac{1}{\pi} \left( \varepsilon_d + \frac{U}{2}
\right) = \int_{-\infty}^Q \sigma_c (\lambda) d \lambda,
\]
where $\Lambda$ and $Q$ denote the \changes{Fermi points} of the unbound
charge and spin excitations respectively. We have $Q=-\infty$ at the
particle-hole symmetric point ($\varepsilon_d = -U/2$) and spin
excitations are absent in the ground state. Similarly, unbound charges
do not exist without external magnetic field and $\Lambda=-\infty$ in
that case.

The impurity spin and charge densities $\sigma_i (\lambda)$ and
$\rho_i (k)$ describe changes in the ground state when the coupling to
the dot is included. They are related to the occupation number $\nd$
and the magnetization $\md=(\ndup-\nddown)/2$ of the dot
through~\cite{kawakami1982ground}
\begin{equation}\label{magnetization}
\md = \frac 1 2 \int_{-\infty}^\Lambda \rho_i (k) d k,
\qquad \nd = 1- \int_{-\infty}^Q \sigma_i (\lambda) d \lambda,
\end{equation}
and we recover the fact that $\nd=1$ at the particle-hole symmetric
point, and $\md=0$ when no magnetic field is applied.

The densities $\sigma_c (\lambda)$, $\rho_c (k)$, 
$\sigma_i (\lambda)$ and $\rho_i (k)$, characterizing the ground state, are solution of the coupled linear integral equations ($a=c/i$)
\begin{subequations}\label{lineareq}
\begin{align}
\begin{split}
\label{lineareq1}
\rho_a (k) + g'(k)  \int_{-\infty}^\Lambda d k' \, R\left[ g(k) - g(k') \right]
\rho_a (k')  \\ + g'(k) \int_{-\infty}^Q d \lambda \, 
s \left[ g(k) - \lambda \right] \sigma_a (\lambda)
 = { \cal S}_{a,1} & (k),
\end{split} \\[2mm]
\label{lineareq2}
\begin{split}
\sigma_a (\lambda) -  \int_{-\infty}^Q d \lambda \, R (\lambda-\lambda') 
\sigma_a(\lambda')  \\ + \int_{-\infty}^\Lambda d k \, s\left[ \lambda - g(k) \right]
\rho_a (k)   = { \cal S}_{a,2} & (\lambda).
\end{split}
\end{align}
\end{subequations}
The conduction and impurity equations differ only by the source term
in the right-hand-side
\begin{align}\nonumber
{ \cal S}_{c,1} (k) &= \frac{1}{2 \pi} \left[
1 + g'(k) \int_{-\infty}^{+\infty} d k' \, R\left[ g(k) - g(k')
\right] \right], \\[2mm]\nonumber
{ \cal S}_{c,2} (\lambda) &= \frac{1}{2 \pi} 
\int_{-\infty}^{+\infty} d k \, s\left[ \lambda - g(k)
\right], \\[2mm]\nonumber
{ \cal S}_{i,1} (k) &= \Delta(k) + g'(k) 
\int_{-\infty}^{+\infty} d k' \, R\left[ g(k) - g(k')
\right] \Delta(k'), \\[2mm]\nonumber
{ \cal S}_{i,2} (\lambda) &=  
\int_{-\infty}^{+\infty} d k \, s\left[ \lambda - g(k)
\right] \Delta(k),
\end{align}
with the definitions
\begin{subequations}
\begin{align}
R(x) & = \frac{1}{2 \pi} \int_{-\infty}^{+\infty} d \omega \, 
\frac{e^{- i \omega x}}{1+e^{|\omega|}}, \\[2mm]
s(x) &= \frac{1}{2 \cosh(\pi x)}, \\[2mm]
\label{functions}
g(k) &= \frac{1}{2 U \Delta} \left(k - \varepsilon_d - U/2
\right)^2, \\[2mm] \Delta(k) & = \frac{1}{\pi}
\frac{\Delta}{(k-\varepsilon_d)^2 + \Delta^2}.
\end{align}
\end{subequations}

\subsection{Wiener-Hopf solution}\label{appen-wiener}

A complete analytical solution to the coupled
equations~\eqref{lineareq} does not exist in the general case, for
which they can be solved numerically. Nevertheless, analytical
progress is possible close to the particle-hole symmetric point, or
for a weak magnetic field, in which cases the two equations decouple.

At zero magnetic field $\Lambda = -\infty$ and the second integral equations simplify to
\begin{equation}
\sigma_a (\lambda) -  \int_{-\infty}^Q d \lambda \, R (\lambda-\lambda') 
\sigma_a(\lambda')   = { \cal S}_{a,2}  (\lambda).
\end{equation}
These two equations are solvable by the Wiener-Hopf technique. Details
on this calculation can be found in the
review~\cite{tsvelick1983}. The result is a parametric expression of
$\nd$ as a function of $\varepsilon_d$ via the \changes{Fermi point} $Q$,
namely
\begin{equation}\label{par1}
\begin{split}
\varepsilon_d & = -\frac{U}{2} + \sqrt{2 U \Delta Q} \theta(Q) - \frac{\sqrt{U \Delta}}{2 \pi^{3/2}}  {\rm Re} \Bigg[ \frac{1}{\sqrt{i}} \int_0^{+\infty}d \omega \\ 
& \times\frac{e^{-2 i Q \pi \omega}}{\omega^{3/2}}  \left \{ e^{-\pi \omega} \left( \frac{e}{i \omega} \right)^{i \omega} \Gamma\left( \frac{1}{2} + i \omega \right) -\sqrt{\pi} \right \} \Bigg],
\end{split}
\end{equation}
or, alternatively,
\begin{equation}\label{par1al}
\begin{split}
 \varepsilon_d = -\frac{U}{2} & + 2 \sqrt{\frac{U \Delta}{2 \pi}} \sum_{n=0}^{+\infty} \frac{(-1)^n}{n! (1+ 2 n)^{3/2}}  \left(\frac{n+1/2}{e} \right)^{n+1/2} \\
&+ \frac{\sqrt{U \Delta}}{2 \pi^{3/2}}  {\rm Re} \Bigg[ \frac{1}{\sqrt{i}} \int_0^{+\infty}d \omega 
\frac{1 - e^{-2 i Q \pi \omega}}{\omega^{3/2}}  \\ & \times e^{-\pi \omega} \left( \frac{e}{i \omega} \right)^{i \omega} \Gamma\left( \frac{1}{2} + i \omega \right)  \Bigg],
\end{split}
\end{equation}
both valid for all $Q$. $\Gamma(z)$ denotes the gamma function. An alternative summation can be found in Ref.~\cite{tsvelick1983} for $Q<0$ but it does not yield a sizeable numerical speed-up. The second expression is
\begin{equation}\label{par2}
\begin{split}
  & \nd  = \frac{1}{2} - \frac{1}{\pi^{3/2}} {\rm Re} 
\Bigg[  \int_0^{+\infty} i d \omega 
\frac{e^{-2 i Q \pi \omega}}{\omega} e^{-\pi \omega} 
\left( \frac{e}{i \omega} \right)^{i \omega} \\
  & \times \Gamma \left( \frac{1}{2} + i \omega \right)
  \int_{-\infty}^{+\infty} \frac{d x}{\pi} \frac{e^{i \pi \omega x^2
      \Delta/U}}{1+(x+U/2 \Delta)^2} \Bigg].
\end{split}
\end{equation}
Eqs.~\eqref{par1} and~\eqref{par2} can be used to compute $\nd$ and
therefore $\delta_0$. The charge susceptibility $\chi_c$ is obtained
from the derivatives of these two expressions with respect to $Q$, and
\[
\chi_c = - \frac{\partial \nd/\partial Q}{\partial
  \varepsilon_d/\partial Q}.
\]
In order to compute the spin susceptibility, we need to add a small
magnetic field. The two equations~\eqref{lineareq} are then weakly
coupled and can be solved perturbatively at low magnetic
field~\cite{tsvelick1983}. The result for the spin susceptibility at
zero magnetic field is finally given by
\begin{equation}
\chi_s = \frac{e^{\pi Q} \, \bar{\sigma}_i  + e^{\pi/I}  + \int_x \frac{1}{\pi} \frac{e^{- \pi  x^2 \Delta/2 U}}{1+(i x+U/2 \Delta)^2}}{2 \sqrt{2 U \Delta} + 4 \pi \sqrt{U \Delta}  \,  e^{\pi Q}  \, \bar{\sigma}_c },
\end{equation}
where $1/I = U/8 \Delta - \Delta/2 U$, with
\begin{equation}
\begin{split}
\bar{\sigma}_c =  - \frac{1}{2 \pi^2 \sqrt{2 e}} & {\rm Re} \Bigg[  \int_0^{+\infty} d \omega \frac{e^{-2 i Q \pi \omega}}{\omega+i/2} \\ 
& \times \frac{e^{-\pi \omega}}{\sqrt{i \omega}} \left( \frac{e}{i \omega} \right)^{i \omega} \Gamma\left( \frac{1}{2} + i \omega \right) \Bigg],
\end{split}
\end{equation}
and
\begin{equation}
\begin{split}
 \bar{\sigma}_i & = \frac{1}{\pi \sqrt{2 e}} {\rm Re} \Bigg[  \int_0^{+\infty} i d \omega \frac{e^{-2 i Q \pi \omega}}{\omega+i/2} e^{-\pi \omega} \left( \frac{e}{i \omega} \right)^{i \omega} \\ 
& \times  \Gamma \left( \frac{1}{2} + i \omega \right) 
\int_{-\infty}^{+\infty} \frac{d x}{\pi} \frac{e^{i \pi \omega x^2 \Delta/U}}{1+(x+U/2 \Delta)^2} \Bigg].
\end{split}
\end{equation}

\subsection{Mixed-valence regime}\label{appen:mixed}

The Bethe Ansatz solutions derived in Sec.~\ref{appen-wiener} for
$\nd$ and $\chi_s$ simplify substantially in the mixed-valence limit
where $U \to \infty$ with fixed $\varepsilon_d$ and $\Delta$. In this
limit, the \changes{Fermi point} $Q$ becomes very large. It can be
absorbed into the definition of a renormalized single-particle energy
\begin{equation}\label{renorm}
2 \Delta Q - \frac{U}{4} = \varepsilon_{dR} = \varepsilon_d + \frac{\Delta}{\pi} \ln \left( \frac{\pi {\rm e} U}{4 \Delta} \right).
\end{equation}
This result is obtained because we took the limit of large $U$ after taking the limit of an infinite cutoff for the Anderson model. If the opposite is done, the same theory applies by with the model high-energy cutoff (bandwidth) replacing $U$ in Eq.~\eqref{renorm}. All observables are now universal functions of $\varepsilon_{dR}$ and $\Delta$, namely the dot occupancy is given by
\begin{equation}
\begin{split}
\nd = \frac{1}{2} - \frac{1}{\pi^{3/2}} \int_0^{+\infty} d \omega \, e^{-2 \pi \omega} \,
{\rm Re} \bigg[i \frac{e^{-i \pi \omega \varepsilon_{dR}/\Delta}}{\omega} \\
\times  \Gamma \left( \frac{1}{2}
+ i \omega \right)  \left( \frac{e}{i \omega} \right)^{i \omega} \bigg].
\end{split}
\end{equation}
This expression is suitable for fast numerical calculation thanks to its exponential convergence. It is also easy to differentiate, the charge susceptibility then takes the form 
\begin{equation}
\label{mixed-chic}
\begin{split}
\chi_c = \frac{1}{\sqrt{\pi} \Delta} \int_0^{+\infty} d \omega \, e^{-2 \pi \omega} \,
{\rm Re} \bigg[ e^{-i \pi \omega \varepsilon_{dR}/\Delta} \, \\
\times  \Gamma \left( \frac{1}{2}
+ i \omega \right)  \left( \frac{e}{i \omega} \right)^{i \omega} \bigg].
\end{split}
\end{equation}
The spin susceptibility also simplifies to
\begin{equation}\label{mixed-chis}
\begin{split}
\chi_s = \frac{\sqrt{2 \pi e}}{8 \, \Delta}  \, e^{-\pi \varepsilon_{dR}/(2 \Delta)} + 
\frac{1}{8 \sqrt{\pi} \, \Delta} 
\int_0^{+\infty} d \omega \\[2mm]  e^{-2 \pi \omega} 
{\rm Re} \left[i \frac{e^{-i \pi \omega \varepsilon_{dR}/\Delta}}{\omega+i/2} \, \Gamma \left( \frac{1}{2}
+ i \omega \right)  \left( \frac{e}{i \omega} \right)^{i \omega} \right].
\end{split}
\end{equation}

\section{Conformal field theory justification of the low-energy
Hamiltonian}
\label{appen-cft}

The structure of the low energy model (see Eq.~\eqref{eq:H_FL})
\begin{eqnarray}
\HFL & =&   \sum_\sigma \int_\varepsilon \, ( \varepsilon -\sigma B/2) \; b_{\varepsilon \sigma}^\dagger b_{\varepsilon \sigma} +H_\alpha + H_\phi +\dots
\label{eq:H_FL2}
\\
H_\alpha&=& - \! \sum_\sigma \int_{\varepsilon_{1},\varepsilon_2} 
\!\!
 \left[ \frac{\alpha_1}{2\pi}  \bigl({\varepsilon_1 + \varepsilon_2}\bigr) + 
 \frac{\alpha_2}{4\pi}\bigl({\varepsilon_1 + \varepsilon_2}\bigr)^2 \right]  
\!  b_{\varepsilon_1 \sigma}^\dagger b_{\varepsilon_2 \sigma}
 \nonumber
\\
\nonumber
H_\phi&=&
  \int_{\varepsilon_{1},\dots,\varepsilon_4} 
\left[ \frac{\phi_1}{\pi} + \frac{\phi_2}{4\pi} 
(\sum_{i=1}^4 \varepsilon_i)  
\right]
: b_{\varepsilon_1 \uparrow}^\dagger  b_{\varepsilon_2 \uparrow}  b_{\varepsilon_3 \downarrow}^\dagger  b_{\varepsilon_4 \downarrow}  :,
\end{eqnarray}
  can be justified by adapting conformal field theory arguments,
  formulated by Affleck and Ludwig \cite{affleck1991b,affleck1993} and
  Lesage and Saleur \cite{lesage1999a,*lesage1999b} 
  in the context of the Kondo model, to the present case of the
  Anderson model. The only difference is that the Anderson model lacks
  the particle-hole symmetry possesed by the Kondo model, thus it has
  more operators perturbing the IR fixed point.

The infrared fixed point is described by the conformally invariant
action~
\begin{equation}\label{infra}
S_0 = \sum_\sigma \int_0^\beta d \tau \int_{-\infty}^{\infty} d x \, b_{x,\tau \sigma}^\dagger \left( \partial_\tau - i v_F \partial_x \right) b_{x,\tau \sigma}
\end{equation}
where the chiral left-moving field 
\begin{equation}\label{leftmov}
b_{x,\tau \sigma} = \int_{-\infty}^{\infty} \frac{d \varepsilon}{\sqrt{2 \pi v_F}} \, e^{i \varepsilon x/v_F} b_{\varepsilon \tau \sigma}
\end{equation}
is a function of $z \equiv \tau + i x$ only. Hence, it satisfies the holomorphic property
\begin{equation}\label{holomorphic}
\partial_\tau b_{x,\tau \sigma} = - i \partial_x b_{x,\tau \sigma}.
\end{equation}
We note that each derivative of Eq.~\eqref{leftmov} with respect to
$x$ produces an additional power of the energy, $\partial_x b_{0,\tau
  \sigma} \leftrightarrow \varepsilon b_{\varepsilon \tau \sigma}$.

At low energy, the infrared action Eq.~\eqref{infra} is complemented
by irrelevant operators. These operators can be constructed quite
generally using the following rules: (i) they are normal ordered
products of $b^\dagger$ and $b$ operators (same number of each), or
derivatives thereof, (ii) they must respect the SU(2)-spin symmetry of
the original Anderson model and conserve spin, (iii) all fields are
taken at $x=0$. Rule (iii) removes automatically all combinations
where the same operator appears twice, for instance $b_\sigma
b_\sigma=0$, as a result of the Pauli principle. Although these rules
allow for an infinite number of terms, operators can be classified
according to their dimension. Adding a pair of fields $b$ and
$b^\dagger$ or taking one derivative with respect to $x$ increases the
dimension by one. An operator with dimension $n$ gives, to leading
order, an energy correction $\propto \varepsilon^{n-1}$ where
$\varepsilon$ can be $V$, $B$ or $T$. In this work, we keep only the
leading and sub-leading irrelevant operators of dimension $2$
($\alpha_1$ and $\phi_1$) and $3$ ($\alpha_2$ and $\phi_2$).

There is an additional simplification to this problem, namely the
equivalence of two operators which differ by a total derivative. Let
$A_0$ denote an even product of $b,b^\dagger$ or derivatives of
$b,b^\dagger$ fields taken at $x=0$. Then, Eq.~\eqref{holomorphic}
implies that its contribution to the action,
\begin{equation}
\begin{split}
\int_0^\beta \, \partial_x A_0 (\tau) & = -i \int_0^\beta \, \partial_\tau A_0 (\tau) \\
& = i \left( A_0(0)  - A_0 (\beta) \right) = 0,
\end{split}
\end{equation}
vanishes due to the antiperiodic boundary conditions in time for
fermions. For example, $(- \partial_x b_{0,\tau \sigma}^\dagger)
b_{0,\tau \sigma} \leftrightarrow \varepsilon_1 b_{\varepsilon_1,\tau
  \sigma}^\dagger b_{\varepsilon_2,\tau \sigma}$ and $ b_{0,\tau
  \sigma}^\dagger (\partial_x b_{0,\tau \sigma}) \leftrightarrow
\varepsilon_2 b_{\varepsilon_1,\tau \sigma}^\dagger
b_{\varepsilon_2,\tau \sigma}$ are equivalent since their difference
is a total derivative $\partial_x (b_{0,\tau \sigma}^\dagger b_{0,\tau
  \sigma})$.

Now, let us classify the possible operators. There is a single
dimension $1$ (marginal) operator $\sum_\sigma b_{0,\tau
  \sigma}^\dagger b_{0,\tau \sigma}$, corresponding to potential
scattering. By a change of basis, it can be absorbed into the action
Eq.~\eqref{infra} where it tunes the zero-energy phase shift
$\delta_0$. Dimension $2$ is obtained by adding a pair of $b$ and
$b^\dagger$ fields. The only possibility fulfilling the conditions
(i), (ii) and (iii) is given by $b_{0,\tau \uparrow}^\dagger b_{0,\tau
  \downarrow}^\dagger b_{0,\tau \downarrow} b_{0,\tau \uparrow}$,
corresponding, after using the decomposition Eq.~\eqref{leftmov} over
energies, to the $\phi_1$ term in the Hamiltonian
Eq.~\eqref{eq:H_FL2}. Dimension $2$ is also obtained from
  $b^\dagger b$ by applying a derivative $\partial_x$ to either $b$
or $b^\dagger$. The two options are nonetheless equivalent, as noted
above, because they differ by a total derivative. After going to
energy space with Eq.~\eqref{leftmov}, one obtains the $\alpha_1$ term
in the Hamiltonian Eq.~\eqref{eq:H_FL2}. The symmetric writing with
respect to $\varepsilon_{1/2}$ in Eq.~\eqref{eq:H_FL2} has been chosen
for aesthetic reasons but any non-symmetric combination of
$\varepsilon_{1/2}$ would also be correct.

We turn to dimension $3$ operators. Point (iii) with the Pauli
principle excludes the choice of six fields. They can not all be
different as we have at most two spin species, in contrast to
situations with higher spin representations considered
in~\cite{mora2009a}. However four fields with a spatial derivative
$\partial_x$ is possible. Applying $\partial_x$ to the $b_\sigma$ in
an SU(2) symmetric way is, up to a total derivative, the same as
applying $\partial_x$ to the $b_\sigma^\dagger$, \changes{and both are
  equivalent to multiplication by a factor of energy}. Therefore,
there is a single operator, one possible writing being the $\phi_2$
term in the Hamiltonian Eq.~\eqref{eq:H_FL2}. The last option for
dimension $3$ is to have two fields and two spatial derivatives. Up to
total derivatives, the energy dependences $\varepsilon_1^2$,
$\varepsilon_1 \varepsilon_2$ and $\varepsilon_2^2$ in front of
$b_{\varepsilon_1,\tau \sigma}^\dagger b_{\varepsilon_2,\tau \sigma}$
are in fact equivalent. Therefore, there is again a unique
inequivalent operator, given by the $\alpha_2$ term in the Hamiltonian
Eq.~\eqref{eq:H_FL2}, the choice of prefactor
$(\varepsilon_1+\varepsilon_2)^2$ being arbitrary.

To summarize,  the above arguments imply
that the holomorphic property of the infrared field and the SU(2)-spin
symmetry constrain the low energy model to the form of the Hamiltonian
Eq.~\eqref{eq:H_FL}.

\section{T-matrix expression}\label{appen-self}

In cases where the Anderson model describes an impurity in a metallic
host, it is instructive to compute the T-matrix ${\cal T}_\sigma
(\varepsilon)$ which characterizes scattering of conduction electrons
by the localized impurity. It is defined through
\begin{equation}
\begin{split}
{\cal G}_{\sigma,{\bf k},{\bf k'}} (\varepsilon) = &{\cal G}^0_{\sigma,{\bf k}} (\varepsilon) \delta({\bf k}-{\bf k'}) \\[1mm]
&+ {\cal G}^0_{\sigma,{\bf k}} (\varepsilon) {\cal T}_\sigma (\varepsilon)
{\cal G}^0_{\sigma,{\bf k'}} (\varepsilon),
\end{split}
\end{equation}
where ${\cal G}_{\sigma,{\bf k},{\bf k'}}$ and ${\cal
  G}^0_{\sigma,{\bf k}}$ are the full and bare conduction electron
Green's functions respectively (for more details, see
Refs.~\cite{hewson1997kondo,affleck1993,hanl2014}).

The elastic contribution to the self-energy is simply fixed by the phase shift (Eq.~eqref{pshift3} in the main text)
\begin{equation}\label{pshift4}
\delta_\sigma (\varepsilon)
=  \delta_{0}   + \alpha_{1} \varepsilon + \alpha_{2}  \varepsilon^2 
 -  \phi_2 \left( \frac{(\pi T)^2}{12} + \frac{(e V)^2}{16} \right),
\end{equation}
through~\cite{affleck1993}
\begin{equation}\label{self1}
 {\cal T}^{\rm el}_\sigma (\varepsilon) = - \frac{i}{2 \pi \nu_0} \left( 1 - e^{2 i \delta_\sigma (\varepsilon)} \right),
\end{equation}
where $\nu_0$ is the bulk density of states per spin. Recalling that the phase shift $\delta_\sigma$ already includes all Hartree diagrams, one realizes that the leading inelastic contribution to the T-matrix is $\propto \phi_1$, {\it i.e.} to second order in the $\phi_1$ term of the Fermi-liquid Hamiltonian Eq.~\eqref{eq:H_FL2}. This contribution has already been calculated by Affleck and Ludwig~\cite{affleck1993}, with the result
\begin{equation}\label{inel}
{\cal T}^{\rm inel}_\sigma (\varepsilon) = - \frac{i e^{2 i \delta_0}}{2 \pi \nu_0} \phi_1^2 \left[ \varepsilon^2 + (\pi T)^2 \right].
\end{equation}
 To second order in energy $\varepsilon$ and temperature $T$, the full T-matrix,  $ {\cal T}_\sigma = {\cal T}^{\rm el}_\sigma + {\cal T}^{\rm inel}_\sigma$ is obtained by expanding Eq.~\eqref{self1}, after inserting Eq.~\eqref{pshift4},
and adding Eq.~\eqref{inel}. The imaginary part, or local spectral function, takes the form
\begin{equation}
\begin{split}
  - \pi \nu_0 {\rm Im} \, T_\sigma (\varepsilon) = \frac{1}{2} \left( 1- \cos \theta_0 \right) + \alpha_1 \sin ( 2 \delta_0 ) \, \varepsilon \\[2mm]
  + \left[ \cos( 2 \delta_0 ) \left(\alpha_1^2+\tfrac{1}{2}\phi_1^2  \right) + \sin (2 \delta_0) \alpha_2 \right] \varepsilon^2 \\[2mm]
  + \tfrac{1}{2}\left[ \cos( 2 \delta_0) \phi_1^2 - \sin(2 \delta_0)
    \tfrac{1}{6}\phi_2  \right] (\pi T)^2.
\end{split}
\end{equation}
Breaking particle-hole symmetry, $\delta_0 \ne \pi/2$, leads to a
linear energy dependence $\propto \alpha_1 \sin ( 2 \delta_0 )$, in
contrast to the Kondo model. As expected for a spectral function, the
temperature dependence is only due to interactions, $\propto
\phi_1,\phi_2$. Using the results from
  Appendix~\ref{appen-empty} in the main text, one can establish the following points:
  The temperature correction remains negative for all values of $U$
  and $\varepsilon_d$, corresponding to a transfer of spectral weight
  to higher energies. In contrast to that, the $\varepsilon^2$
  coefficient, or spectral function curvature, is negative in the
  Kondo regime but changes sign in the mixed-valence regime. In the
  empty orbital regime, the temperature correction is at most $-\Delta^3/(\pi
  \varepsilon_d^5)$, that is much smaller than the $\varepsilon^2$ coefficient
  asymptotically given by $3 \Delta^2/\varepsilon_d^4$.


%

\end{document}